\RequirePackage{fix-cm}
\documentclass[smallextended]{svjour3}       
\smartqed  
\usepackage{graphicx}
\usepackage{subfigure}
\usepackage[authoryear]{natbib}
\usepackage{hyperref}
\usepackage{amssymb}

\begin{document}

\title{On the optimal calibration of VVV photometry
}


\author{Gergely~Hajdu    \and
        Istv\'an~D\'ek\'any    \and
        M\'arcio~Catelan    \and
        Eva~K.~Grebel
}


\institute{G. Hajdu \at
              Instituto de Astrof\'isica, Facultad de F\'isica,
              Pontificia Universidad Cat\'olica de Chile, \\
              Av. Vicu\~na Mackenna 4860,
              7820436 Macul, Santiago, Chile \\
              Millennium Institute of Astrophysics, Santiago, Chile\\
              Nicolaus Copernicus Astronomical Center, Polish Academy of Sciences\\
              Bartycka 18, 00-716 Warszawa, Poland\\
              \email{ghajdu@camk.edu.pl}
              \and
              I. D\'ek\'any \at
              Astronomisches Rechen-Institut, Zentrum f\"ur Astronomie der Universit\"at Heidelberg, \\
              M\"onchhofstr. 12-14, 69120 Heidelberg, Germany \\
              \email{idekany@uni-heidelberg.de}
              \and
              M. Catelan \at
              Instituto de Astrof\'isica, Facultad de F\'isica,
              Pontificia Universidad Cat\'olica de Chile, \\
              Av. Vicu\~na Mackenna 4860,
              7820436 Macul, Santiago, Chile \\
              Millennium Institute of Astrophysics, Santiago, Chile\\
              Centro de Astro-Ingenier\'ia,
              Pontificia Universidad Cat\'olica de Chile, \\
              Av. Vicu\~na Mackenna 4860,
              7820436 Macul, Santiago, Chile \\
              \email{mcatelan@astro.puc.cl}
              \and
              E.~K. Grebel \at
              Astronomisches Rechen-Institut, Zentrum f\"ur Astronomie der Universit\"at Heidelberg, \\
              M\"onchhofstr. 12-14, 69120 Heidelberg, Germany \\
              \email{grebel@ari.uni-heidelberg.de}
}

\date{Received: date / Accepted: date}

\maketitle

\begin{abstract}

Prompted by some inconsistencies in the photometry of the VISTA Variables in the V\'ia L\'actea (VVV) survey,
we conduct a revision of the standard calibration procedure of VISTA data in the $J$, $H$, and $K_S$ passbands.
Two independent sources of bias in the photometric zero-points are identified:
First, high sky backgrounds severely affect the $H$-band measurements, but this can mostly be minimized
by strict data vetting. Secondly, during the zero-point calibration, stars serving as standards are taken from
the 2MASS catalog, which can suffer from high degrees of blending in regions of high stellar density,
affecting both the absolute photometric calibration, as well as the scatter of repeated observations.
The former affects studies that rely on an absolute magnitude scale, while the latter can also affect the shapes
and amount of scatter in the VVV light curves, thus potentially hampering their proper classification.
We show that these errors can be effectively eliminated by relatively simple modifications of the standard calibration procedure,
and demonstrate the effect of the recalibration on the VVV survey's data quality.
We give recommendations for future improvements of the pipeline calibration of VISTA photometry,
while also providing preliminary corrections to the VVV $JHK_\mathrm{S}$ observations as a temporary measure.

\end{abstract}

\section{Introduction}\label{sec:intro}

VISTA Variables in the V\'ia L\'actea \citep[VVV;][]{2010NewA...15..433M} is an ESO Public Survey of the bulge and
an adjoining part of the Galactic mid-plane, covering approximately 520 square degrees of the sky.
It uses the Visible and Infrared Survey Telescope for Astronomy \citep[VISTA;][]{2015A&A...575A..25S},
equipped with the VISTA IR Camera (VIRCAM).
VVV acquired time-series photometry of nearly a billion point sources in the near-infrared $K_S$ band between 2009 and 2015,
and additional sparse photometry in the $Z$, $Y$, $J$, and $H$ bands over its entire area.
An extension of the survey called VVV eXtended \citep[VVVX;][]{2018ASSP...51...63M} is currently in progress.

Standard data products for the VVV survey, as well as other ESO surveys carried out by VISTA, are produced
by the VISTA Data Flow System \citep[VDFS,][]{2004SPIE.5493..401E,2004SPIE.5493..411I}.
It comprises a series of software pipelines for quality control and system monitoring by ESO,
image processing, source detection, photometry and calibration by the Cambridge Astronomy Survey Unit
\citep[CASU, see][]{2009MNRAS.394..675H,2018MNRAS.474.5459G} and data curation by the VISTA Science Archive \citep[VSA,][]{2012A&A...548A.119C}.

The VDFS performs photometry on two different image end products known as {\em pawrints} and {\em tiles}.
In the case of the VVV survey observations discussed here, a pawprint is a single detector frame stack combined from two slightly jittered VIRCAM exposures,
and is thus composed of 16 sub-images, one from each of VIRCAM's detector chips, covering a discontiguous area of the sky.
At each observational {\em epoch}, a series of 6 pawprints are taken over a {\em field} with overlapping offset
pointings in a $\sim$3-minute time interval, in order to fill the gaps between the detector's chips.
Consequently, most of the sources have multiple pawprint-based photometric measurements at each epoch.
Pawprints are further combined into contiguous mosaic images known as {\em tiles}, which lead to a separate
line of photometric data products. Tile-based photometry is less precise and accurate due to the complicated
image processing involved in image mosaicking, but it has a slightly higher limiting magnitude.
In this study, we only consider pawprint-based photometry.
Both photometric data products are calibrated to local secondary standard stars from the 2-Micron All-Sky Survey
\citep[2MASS;][]{2006AJ....131.1163S}.
For a more comprehensive technical description of the data processing by CASU, we refer to
\citet[][and references therein]{2018MNRAS.474.5459G}.
Standard data products from CASU are considered to be science-ready and serve as direct input for the VSA.

The VVV survey's standard data products, as provided by CASU, already formed the basis of hundreds of
scientific publications in diverse interlinked astronomical fields, ranging from studies of the stellar
populations of the Galactic bulge \citep[e.g.,][]{2012A&A...543A..13G,2013ApJ...776L..19D,2013MNRAS.435.1874W},
to analyses of interstellar extinction \citep{2012A&A...543A..13G,2014A&A...566A.120S,2016MNRAS.456.2692N}
and the census and characterization of stellar clusters in the Milky Way
\citep{2011A&A...532A.131B,2012A&A...545A..54C,2015AJ....149...99A}, to mention only a few.
Most of these studies concern objects located toward crowded stellar fields at low Galactic latitudes highly
reddened by interstellar dust, where VVV has a unique advantage over optical surveys.
The accuracy of the absolute calibration of VVV photometry under such adverse observational conditions is of
critical importance for the astrophysical characterization of these objects.

Evidence has been accumulating that the standard photometric data products of the VVV survey show significant anomalies.
In this study, we review the pipeline calibration procedure of VISTA photometry by the VDFS, revealing that
under some circumstances,
the photometric zero-points in the $J,H,K_s$ bands are affected by significant time-varying biases
that can adversely affect scientific conclusions, and that can be largely eliminated by further optimization of the calibration method.

\section{Inconsistencies in the VVV photometry}\label{sec:vvv_2mass}

\begin{figure}%
\centering
\subfigure{%
\includegraphics[width=0.45\textwidth]{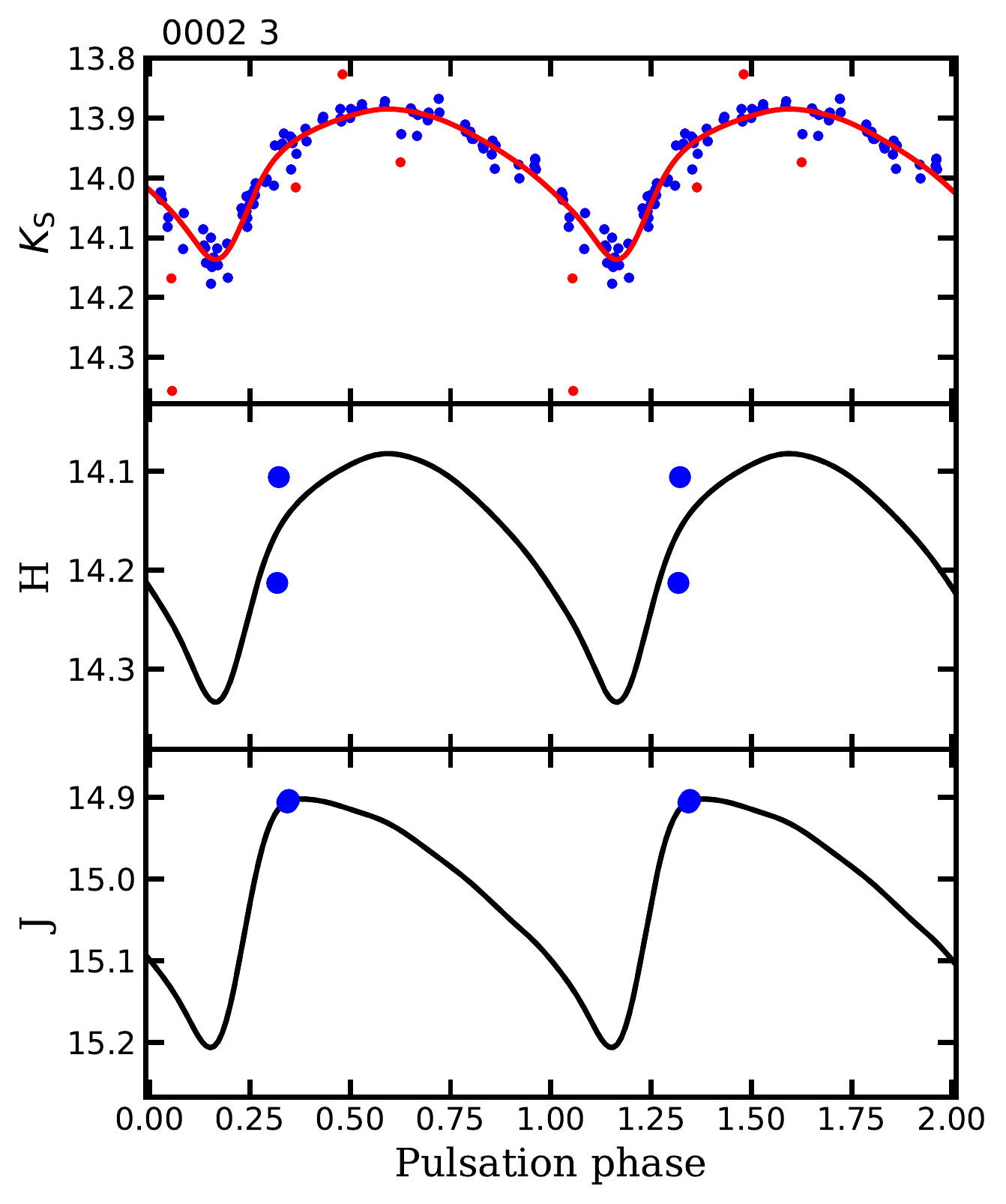}}%
\qquad
\subfigure{%
\includegraphics[width=0.45\textwidth]{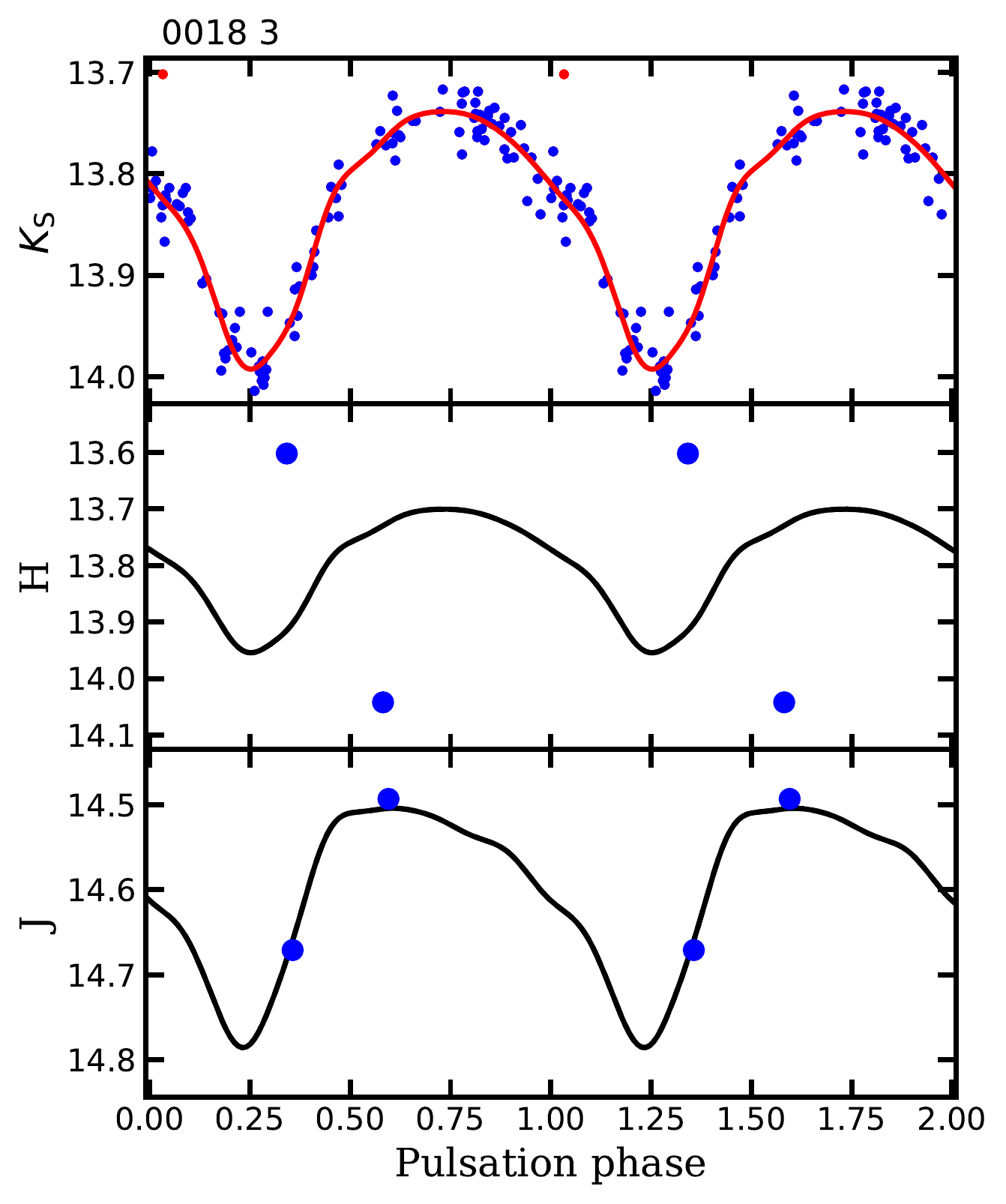}}%
\caption{
Phase-diagrams of two RR~Lyrae stars from \citet{2018ApJ...857...54D} showing their $J,H,K_s$ photometry from the VVV survey. 
The red curves show light curve models fitted directly to the $K_S$ data, while the black curves are light variations predicted by the {\tt PyFiNeR} routine of \citet{2018ApJ...857...55H}.
Note the inconsistent $H$-band measurements (middle panels) for both objects.}
\label{fig:h_incons}
\end{figure}

During our study of the RR~Lyrae stars along the southern Galactic mid-plane \citep{2018ApJ...857...54D}, 
a fraction of our sample showed large inconsistencies between individual $H$-band measurements.
Fig.~\ref{fig:h_incons} shows concrete examples of this for two different stars: on the left panel, the $H$-band measurements of the
star come from two separate pawprints at the same epoch, while on the right panel the $H$-band data come from two different nights.
In both cases, the two $H$-band measurements are inconsistent with the predicted light variation \citep{2018ApJ...857...55H}.
In the study of \citet{2018ApJ...857...54D}, $J$-band measurements did not show such an anomaly, and they were available for almost
all RR~Lyrae stars in our sample, enabling us to estimate their reddening. However, many objects in the VVV survey, such as distant Cepheids,
are extremely obscured by interstellar dust, preventing their detection in the $J$ band. As a result, $H$-band observations
play a crucial role in the characterization of distant, heavily reddened VVV sources. Thus, understanding the cause of the
$H$-band photometric inconsistencies and making the necessary corrections is essential.

An additional source of systematic bias in the VDFS photometry was discovered by \citet{jurcsik18} during their study of
the Blazhko effect in the near-infrared. They note that part of their VVV $K_S$-band RR~Lyrae light curves
showed systematic offsets between measurements corresponding to different pawprints and/or overlapping fields.
These offsets were manually corrected by \citet{jurcsik18} on a relative scale, but this additional source of scatter also
warrants a revision of the VISTA photometry in all passbands.
As the $JHK_\mathrm{S}$ bands are the most relevant for the study of variable stars in the VVV survey,
in this study we have chosen to focus our attention on them.

\section{The calibration of VISTA to 2MASS}\label{sec:vvv_2mass}

The standard procedure of photometric calibration of VISTA data is based on the software pipeline 
by \cite{2009MNRAS.394..675H}, developed for the Wide Field Camera \citep[WFCAM;][]{2007A&A...467..777C}
of the United Kingdom Infrared Telescope (UKIRT) on Mauna Kea, Hawai'i.
CASU have provided multiple versions of the photometry, differentiated by version numbers. 
At the time of writing
{of this paper,
version 1.3 has been completed for the entire VVV data set and version 1.5 is being
deployed.
In the case of the pawprint photometry concerned here, versions
1.3 and 1.5 share the procedures of source extraction and photometry,
with the main differences between the two versions coming from the refined
conversion formulae involved in the absolute calibration \citep{2018MNRAS.474.5459G},
as well as from the modified procedure used to estimate the line-of-sight extinction
towards individual 2MASS stars.
It should be mentioned that further changes were implemented between the two versions
regarding the generation of the tile catalogs, but in this work we are only concerned with the pawprint
photometry.

The VISTA instrumental magnitudes are calibrated to the common VISTA photometric system on a pawprint basis 
by applying a zero-point shift to match the magnitudes of an ensemble of local secondary standard stars 
from the 2MASS survey, converted to the VISTA system.
Changes in the individual detector sensitivities are accounted for by an additional offset, calculated on a monthly basis for every filter.

In order to investigate the previously discussed photometric anomalies, we retrieved the photometry provided in the
2MASS Point Source Catalog \citep{2006AJ....131.1163S} from the Infrared Processing \& Analysis Center\footnote{\url{www.ipac.caltech.edu}}
for all stellar sources in the following areas:

\begin{eqnarray}
\textrm{\bf bulge:} &-10.4^\circ<l<11^\circ \,\, ; \,\, -10.5^\circ<b<5.3^\circ \nonumber \\
\textrm{\bf disk:}  &-65.5^\circ<l<-9^\circ \,\, ; \,\, -2.4^\circ<b<2.4^\circ. \nonumber
\end{eqnarray}

\noindent These two data sets completely encompass the bulge and disk footprints of the VVV survey,
and contain approximately 13.7 and 10.6 million point sources detected by 2MASS, respectively.

Version 1.3 of the CASU photometry uses the following transformation equations between the 2MASS and the VISTA systems \citep{2018MNRAS.474.5459G}:

\begin{eqnarray}
  J_{\mathrm{VISTA}, 1.3}   &=& J_\mathrm{2MASS}     - 0.077 \times (J-H)_\mathrm{2MASS}   + 0.010 \times \mathrm{E}(B-V)_{v1.3} \label{eq:jv13}\\
  H_{\mathrm{VISTA}, 1.3}   &=& H_\mathrm{2MASS}     + 0.032 \times (J-H)_\mathrm{2MASS}   + 0.015 \times \mathrm{E}(B-V)_{v1.3} \\
  K_{S,\mathrm{VISTA}, 1.3} &=& K_{S_\mathrm{2MASS}} + 0.010 \times (J-K_S)_\mathrm{2MASS} + 0.005 \times \mathrm{E}(B-V)_{v1.3}; \label{eq:kv13}
\end{eqnarray}

\noindent while their version 1.5 uses the following equations:

\begin{eqnarray}
  J_{\mathrm{VISTA}, 1.5}   &=& J_\mathrm{2MASS}     - 0.031 \times (J-K_S)_\mathrm{2MASS} + 0 \times \mathrm{E}(B-V)_{v1.5} \label{eq:jv15} \\
  H_{\mathrm{VISTA}, 1.5}   &=& H_\mathrm{2MASS}     + 0.015 \times (J-K_S)_\mathrm{2MASS} + 0 \times \mathrm{E}(B-V)_{v1.5} \label{eq:hv15} \\
  K_{S, \mathrm{VISTA}, 1.5} &=& K_{S_\mathrm{2MASS}} - 0.006 \times (J-K_S)_\mathrm{2MASS} + 0.005 \times \mathrm{E}(B-V)_{v1.5}. \label{eq:kv15}
\end{eqnarray}

\noindent The color excesses $\mathrm{E}(B-V)_{v1.3}$ and $\mathrm{E}(B-V)_{v1.5}$ are estimated differently for each star
in the two versions of the photometry. In both cases, the color excess $\mathrm{E}(B-V)$ is first interpolated at the coordinates of each 2MASS
source from the extinction map of \cite{1998ApJ...500..525S}, originally calculated from the measured infrared emission of dust.
For version 1.3, this $\mathrm{E}(B-V)$ value is modified as:

\begin{eqnarray}
  \mathrm{E'}(B-V) = 0.1 + 0.65 \times (\mathrm{E}(B-V)-0.1),
\end{eqnarray}

\noindent for values bigger than 0.1\,mag, following \cite{2000AJ....120.2065B}. Furthermore, this color excess is limited to
a maximum of 10\,mag, in order to account for its overestimation at low Galactic latitudes. In contrast, for version 1.5, the
interpolated $\mathrm{E}(B-V)$ of \cite{1998ApJ...500..525S} is utilized directly; however, the resulting values are modified to yield a
dereddened $(J-K)_0=0.5$ color for stars where the original $\mathrm{E}(B-V)$ would result in a dereddened color of $(J-K)_0<0.25$.
For the details of the estimations of extinction for 2MASS objects between the two versions of the photometry, as well as of additional
quality cuts applied to the 2MASS sample of point sources, we followed the descriptions of \cite{2018MNRAS.474.5459G}.
These final $\mathrm{E}(B-V)_{v1.3}$ and $\mathrm{E}(B-V)_{v1.5}$ values are used with their corresponding
Eqs.~\ref{eq:jv13}--\ref{eq:kv15} to convert the magnitudes of the 2MASS sources to both versions of the VISTA 
photometric system\footnote{During our discussions with the referee, it has become clear that we have inadvertently
used the modified $\mathrm{E'}(B-V)$ \citep{2000AJ....120.2065B} color excesses in our conversion procedure for
version~1.5 of the VISTA photometric system. As we have otherwise followed the procedures laid out by \cite{2018MNRAS.474.5459G},
this only results in a minor, location-dependent photometric zero-point shift with a maximum value of less than 0.007\,mag
in the $K_\mathrm{S}$ band (in the other bands, the extinction coefficients in the conversion formulae Eqs.~\ref{eq:jv15},\ref{eq:hv15}
are zero). Therefore, we have chosen not to redo our analysis for this study.}.

\cite{2018MNRAS.474.5459G} note that there are small magnitude zero-point offsets between versions 1.3 and 1.5 of
VISTA photometry, mainly caused by the change in the transformation Eqs.~\ref{eq:jv13}--\ref{eq:kv15}. 
Our conversion of the 2MASS photometry to the versions 1.3 and 1.5 of the VISTA system show differences consistent with the ones found by
\citet[][see their Eqs.~C9--C11]{2018MNRAS.474.5459G}, validating our analysis\footnote{\cite{2018MNRAS.474.5459G} calculate these
differences on calibrated VISTA photometry
over different sky regions, while here this was done for the converted 2MASS magnitudes only in the VVV sky
footprint. Hence, these differences are not expected to be exactly the same, owing to the different
sky areas covered by stars of the two samples, as well as the extinction terms in Eqs.~\ref{eq:jv13}--\ref{eq:kv15}.}.

\begin{figure*}[t!]
\centering
\includegraphics[width=\textwidth]{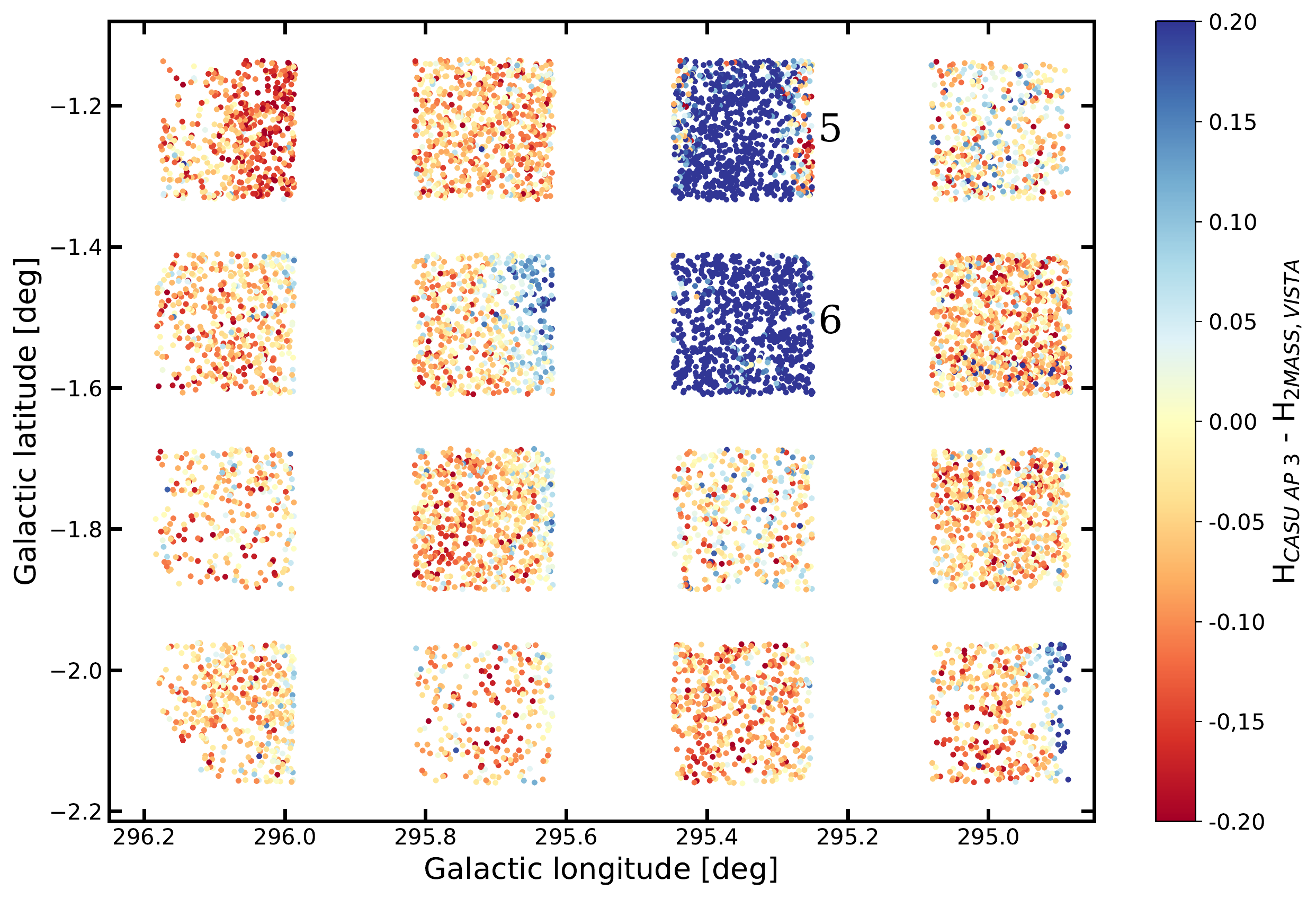}
\caption[$H$-band photometric difference map of a VISTA pawprint]{Photometric difference between VISTA $H$-band observations
and converted 2MASS magnitudes for one pawprint. Chips 5 and 6 (marked on the figure by numbers) possess systematically fainter magnitudes than 
all other detectors. Furthermore, some intrachip differences are also evident when comparing the  two sources of photometry.}
\label{fig:h_diff}
\end{figure*}

\section{Revision of the $H$-band VVV photometry} \label{sec:h_band}

\begin{figure*}[ht!]
\centering
\includegraphics[width=\textwidth]{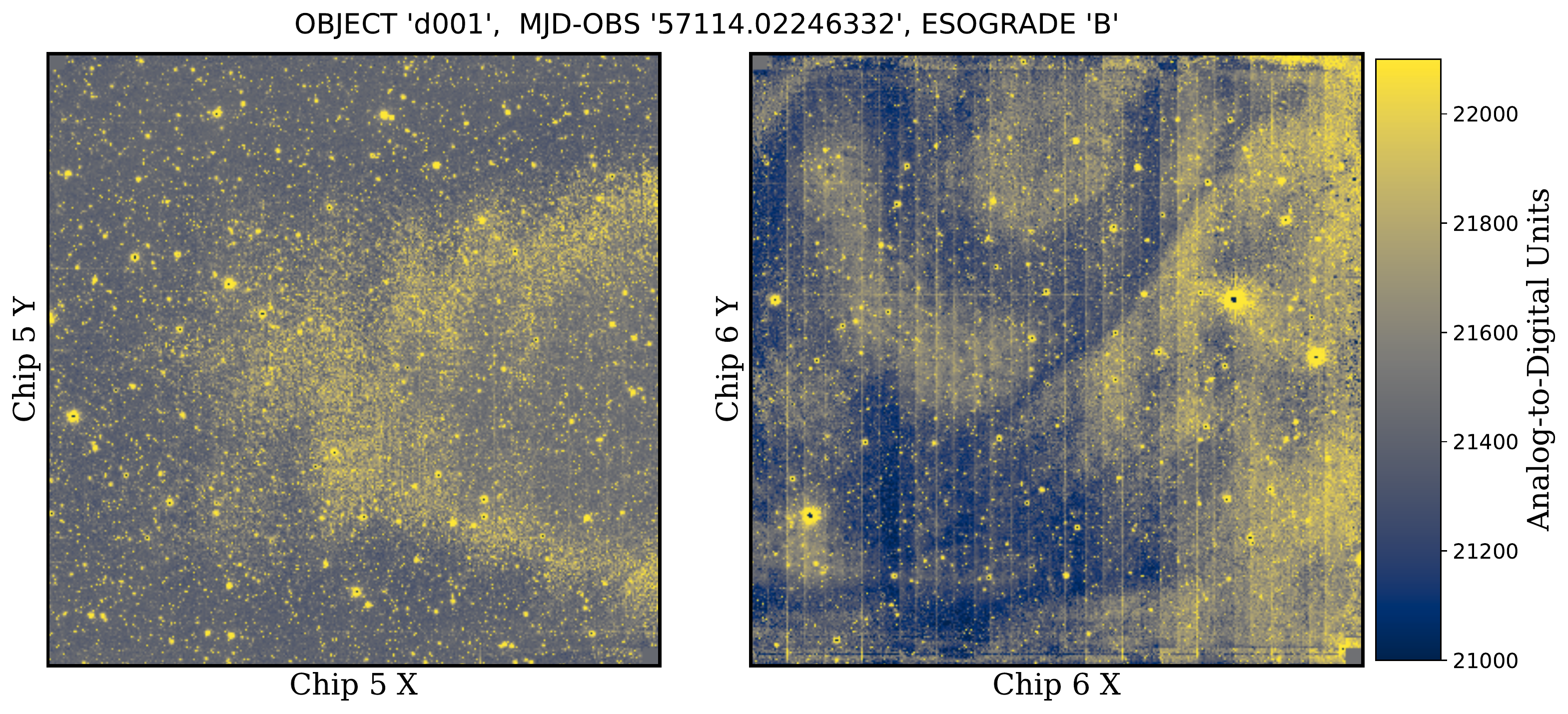}
\caption[Chip 5 and 6 $H$-band images]{The $H$-band (calibrated) images of chips 5 and 6 (marked by numbers), corresponding to the photometric difference
map presented in Fig.~\ref{fig:h_diff}. Both chips suffer from dark regions with diminished point-source analog-to-digital unit counts. The high
background counts indicate suboptimal observing conditions.}
\label{fig:chip5_6}
\end{figure*}

\begin{figure*}[ht!]
\centering
\includegraphics[scale=0.4]{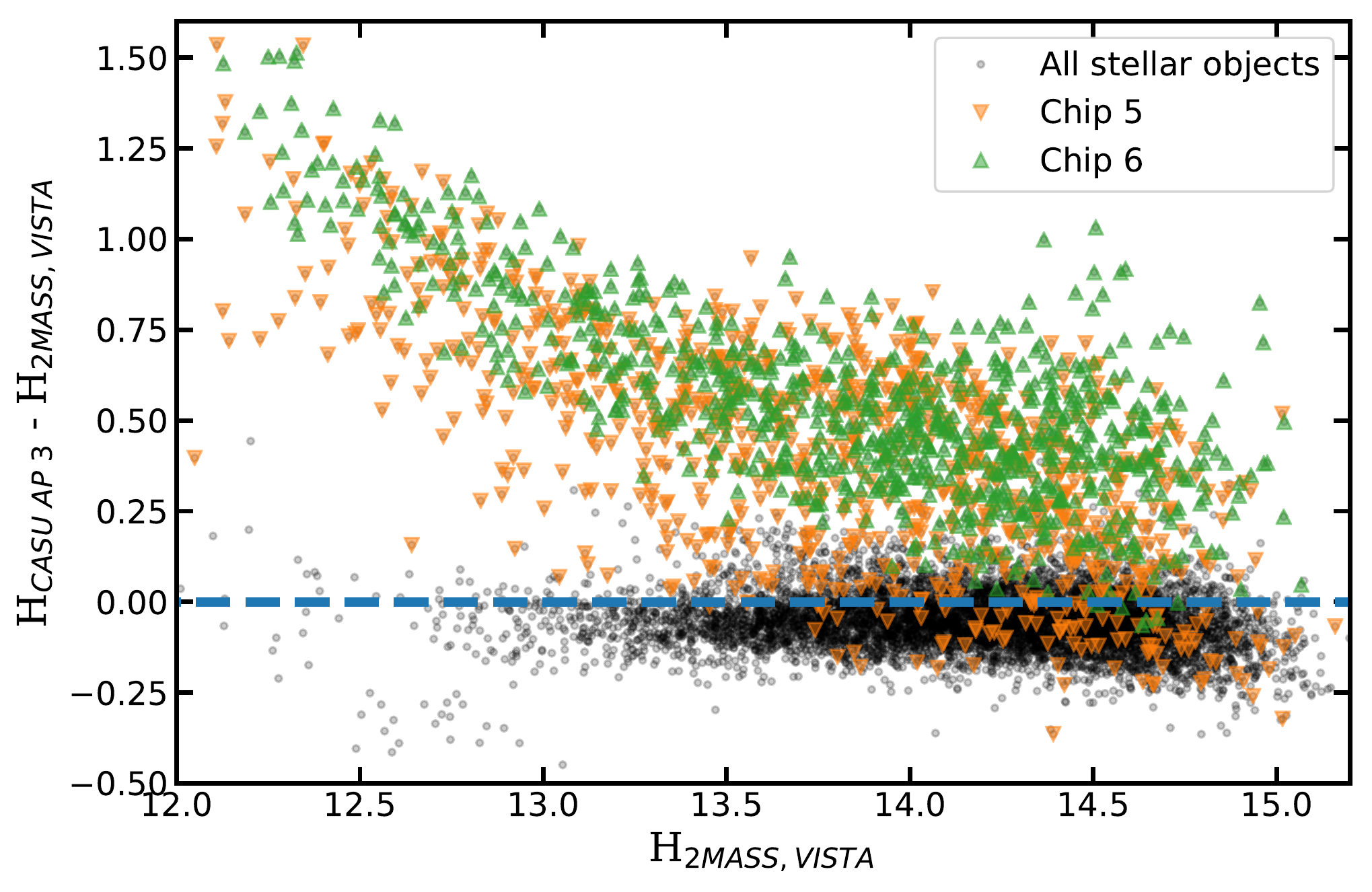}
\caption{$H$-band magnitude differences between the VVV photometry calibrated by CASU and the corresponding converted 2MASS
measurements for the same pawprint as in Figs.~\ref{fig:h_diff} and \ref{fig:chip5_6}. The non-linear response of
the VIRCAM chips 5 and 6 results in lower source counts and biased photometry for these detectors. Furthermore, while the photometry
from the other detectors is mostly internally consistent, the pawprint-wise zero-point calibration scheme employed by CASU results in a systematic overestimation of the point-source magnitudes.}
\label{fig:h_diff2}
\end{figure*}

The converted 2MASS magnitudes obtained in Sect.~\ref{sec:vvv_2mass} allow an independent
check of the photometry delivered by the VDFS.
Fig.~\ref{fig:h_diff} shows the differences between the converted
2MASS $H$-band magnitudes of stellar sources and their VISTA photometry as provided by CASU (version 1.3), 
in one particular pawprint of the low source density VVV field {\em d001} (see \citealt{2010NewA...15..433M} for field definitions).
It is immediately apparent that the joint magnitude zero-point calibration of the entire pawprint causes 
underestimated magnitudes on chips 5 and 6, and generally overestimated magnitudes in the rest of the chips. 
The distribution of magnitude differences even show small-scale systematic variations within some of the chips.

The original, reduced $H$-band VISTA images from chips 5 and 6 are shown in Fig.~\ref{fig:chip5_6}.
Both images exhibit very apparent artifacts, probably caused by the
non-linearity of these two chips when the ``sky background'' (i.e., the atmospheric foreground flux) is high.
The straight lines across chip 6 possibly originate from imperfections in the
detector's manufacturing process.

The distribution of $H$-band magnitude differences for the same pawprint of field {\em d001} are
shown in Fig.~\ref{fig:h_diff2}. In accordance with Fig.~\ref{fig:h_diff}, the magnitudes in chips 5 and 6 are underestimated, while they are generally overestimated on all other chips. Magnitudes measured in
chips 5 and 6 also show large dispersion and non-linearity.
In contrast, the magnitudes from the other chips show
small dispersions and no apparent non-linearity under the same sky conditions.
Therefore, photometric measurements obtained with these chips in the $H$-band should generally be reliable after
a chip-wise zero-point correction.
Meanwhile, $H$-band photometry acquired by chips
5 and 6 under suboptimal conditions should generally be treated as suspect.
Indeed, the same behavior of the VIRCAM detectors can frequently be observed for other 
fields during the entire timespan of the VVV survey.
While it is advisable to avoid using these affected observations, this is not practical in every case.
Unfortunately, for most of the VVV fields, only two $H$-band epochs have been obtained during
the course of the survey, and, depending on the science case, it might be unavoidable to use them.

It has to be noted that similar offsets, apparently caused by unfavorable
observing conditions, were not found in the case of
the $J$ and $K_S$ photometry, as the atmospheric flux in those bands is
generally lower.

\section{VVV magnitude zero-point bias in dense stellar fields} \label{sec:blending}

\begin{figure*}%
\centering
\subfigure{%
\includegraphics[width=0.8\textwidth]{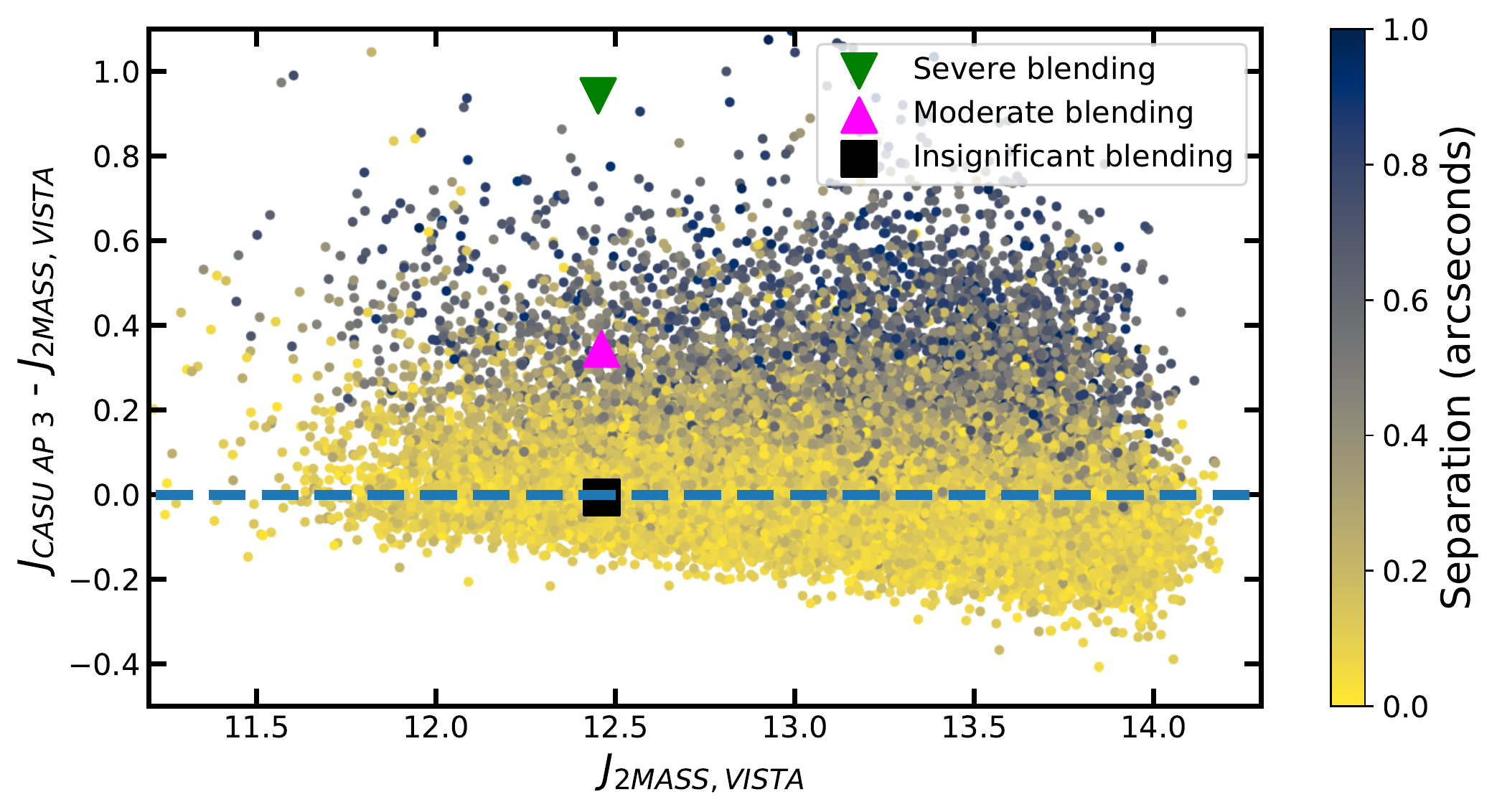}}%
\qquad
\subfigure{%
\includegraphics[width=0.8\textwidth]{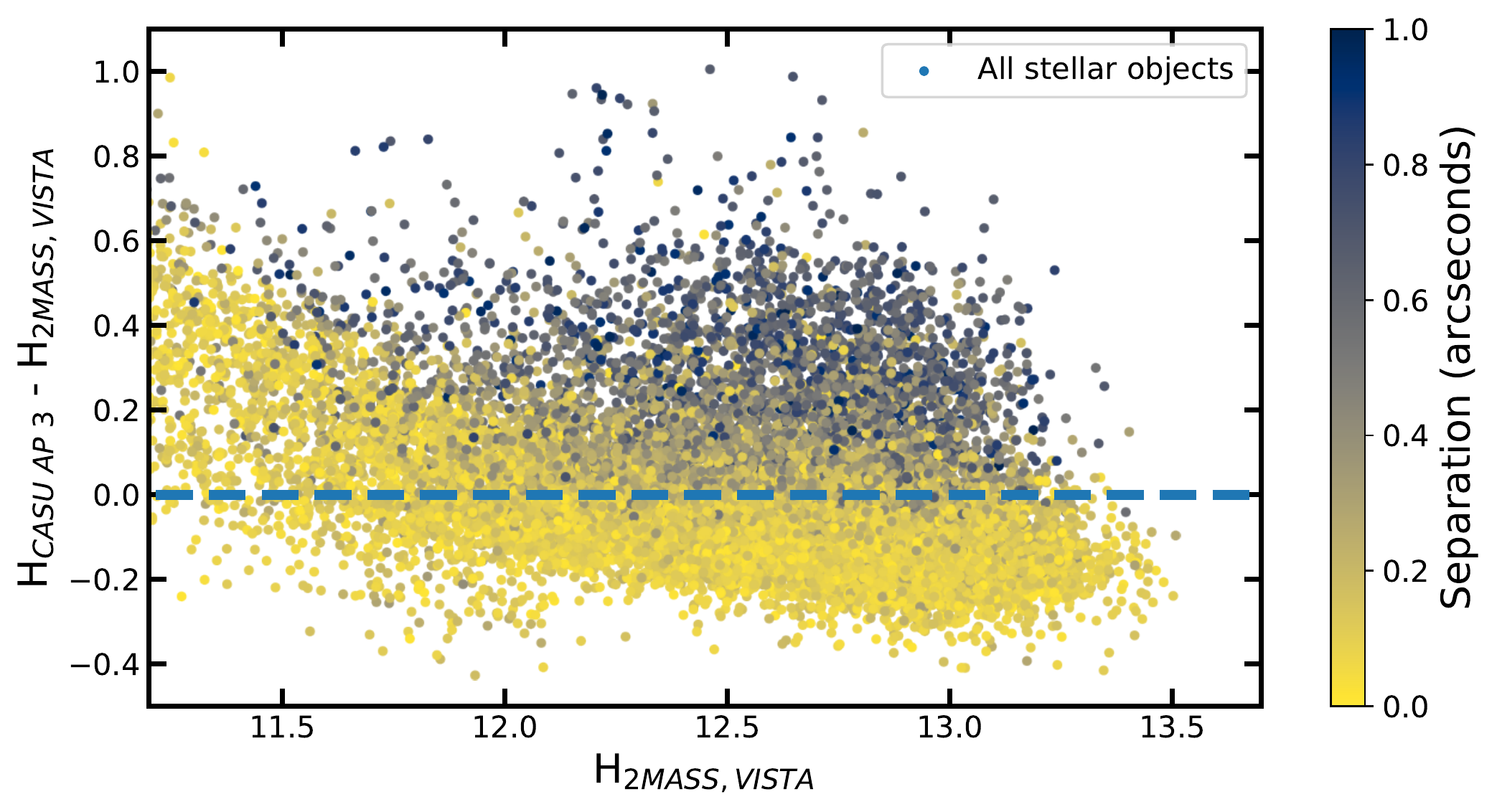}}%
\qquad
\subfigure{%
\includegraphics[width=0.8\textwidth]{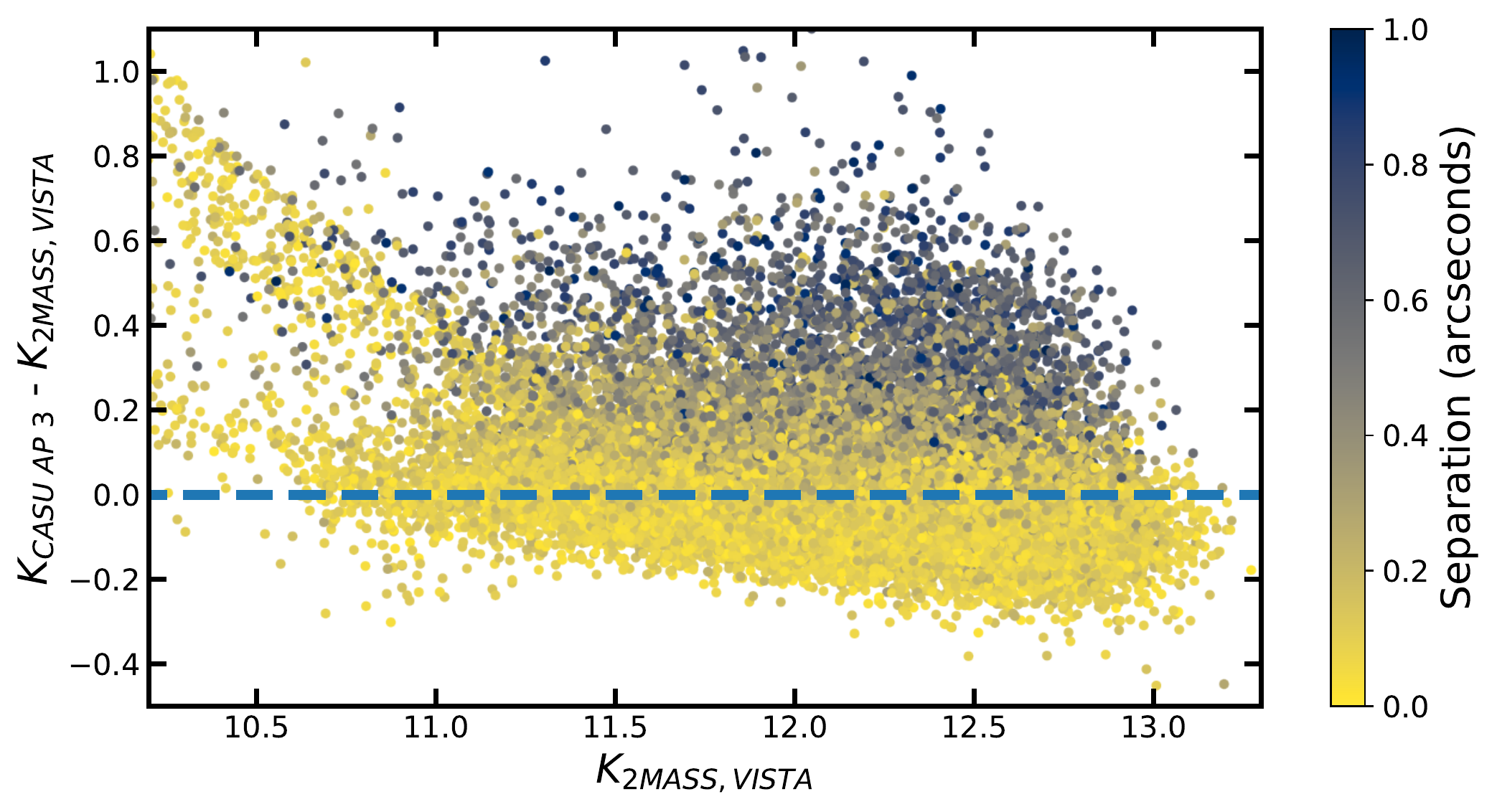}}%
\caption{
$J$, $H$ and $K_\mathrm{S}$-band photometric bias in the VVV field \textit{b307}.
{In all of the panels, 2MASS magnitudes are converted to the VISTA v1.3 system using
formulae Eqs.~\ref{eq:jv13}--\ref{eq:kv13}.
\textit{Top:}
  $J$-band
  magnitude differences in field \textit{b307} between the converted 2MASS measurements and
  the VVV aperture photometry of stellar sources calibrated by CASU. Note the systematic offset from the fiducial line
  for the majority of sources with 2MASS magnitudes fainter than $\sim12$\,mag. The color scale denotes the separation between the
  coordinates of the sources in the 2MASS and VISTA catalogs.
  \textit{Middle:} The same as in the top panel, but for the
  H band. The 2MASS and VISTA images of the three
  individually marked stars are shown in Fig.~\ref{fig:blending}.
  \textit{Bottom:} Same as the other two panels, but for the $K_\mathrm{S}$-band observations.}
}
  \label{fig:b307_jh}
\end{figure*}

The converted 2MASS photometry from Sect.~\ref{sec:vvv_2mass} allows a general revision
of the VISTA photometry for all filters. In the following, we investigate the effect of 
source crowding on the photometric calibration of VVV data in the $J$, $H$, and $K_S$ bands,
in version 1.3 of the VISTA photometric system.

\begin{figure}[]
  \centering
  \includegraphics[width=0.9\textwidth]{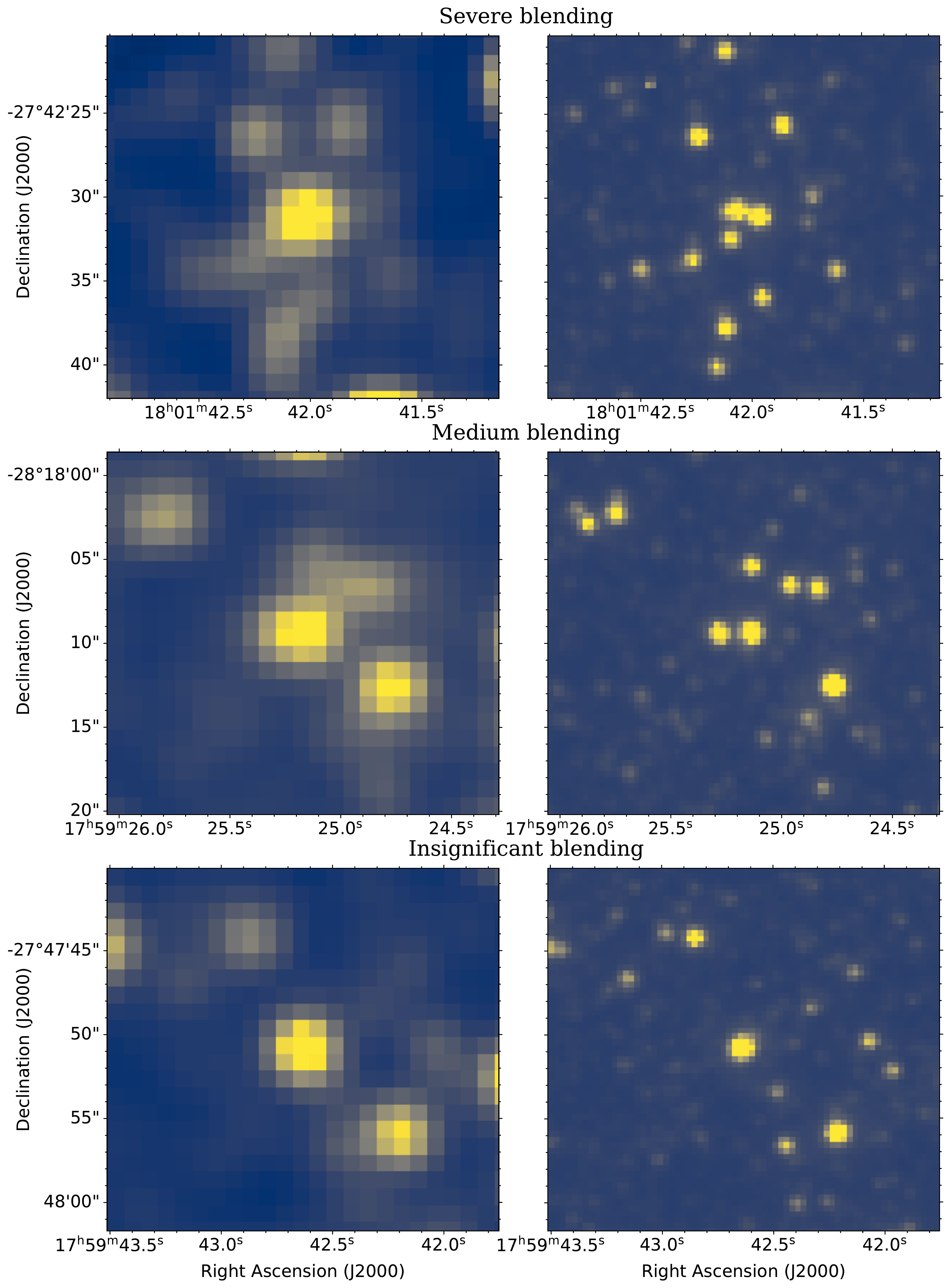}
  \caption{Examples of severe, medium, and insignificant point-source blending in the 2MASS
  $J$-band observations in the area of the VVV survey, shown by the top, middle, and bottom panels, respectively.
  The sources in the image centers correspond to the objects marked by various symbols in the bottom panel 
  of Fig.~\ref{fig:b307_jh}.
  \textit{Left:} $J$-band images of the 2MASS observations. \textit{Right:} $J$-band VISTA images of the same fields.}
\label{fig:blending}
\end{figure}

Aside from the area toward the nuclear bulge, one of the fields with the highest source density is {\em b307}, 
due to its position toward Baade's Window.
The top, middle and bottom panels of Fig.~\ref{fig:b307_jh} compare the 
VDFS-calibrated VVV magnitudes of stellar sources and the
corresponding converted 2MASS measurements for the $J$, $H$, and $K_\mathrm{S}$ bands, respectively,
for all stellar sources from a single pawprint of field {\em b307}.
Most of the fainter stars (2MASS magnitude~$\gtrsim 12$) are offset from the fiducial line.
Furthermore, most outliers from the main locus have positive deviations, meaning that the same sources appear systematically
dimmer in the CASU source catalog with respect to 2MASS. 
Visual inspection of images taken by both surveys revealed that almost all of the latter
stars are blended in the 2MASS observations, while typically being resolved into 2--4 separate point sources by VISTA.
Examples of severe, medium and insignificant blending in the 2MASS observations are shown in Fig.~\ref{fig:blending},
for the individually marked sources in the top panel of Fig.~\ref{fig:b307_jh}.

\begin{figure*}[t!]
  \centering
  \includegraphics[width=0.8\textwidth]{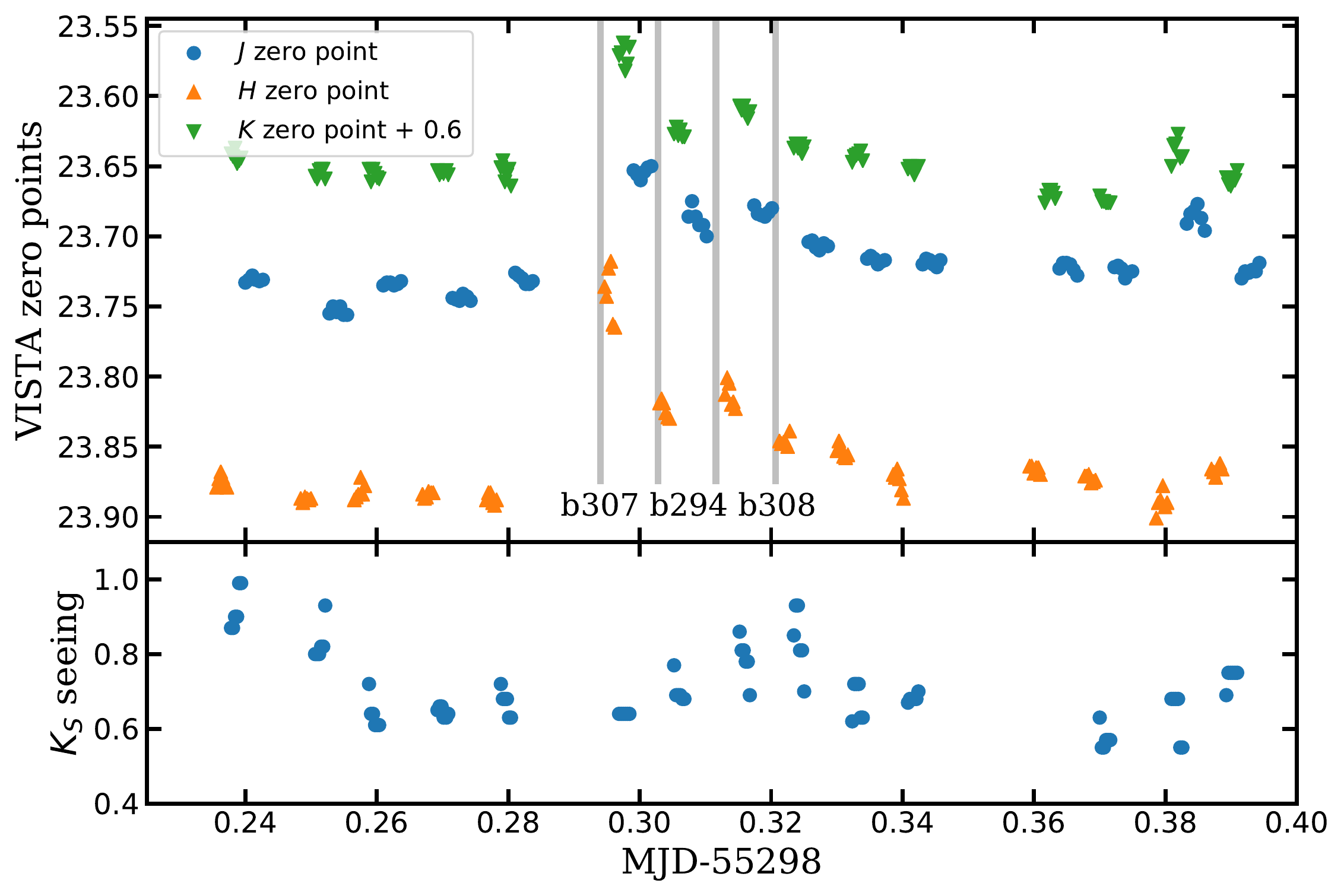}
  \caption[Evolution of the calculated zero-points of VISTA]{\textit{Top:} Evolution of the VIRCAM@VISTA zero-points
  as calculated by the
  CASU pipeline during a period of $4.2$ hours of one night. Zero-points in tiles with high stellar density (especially tiles
  \textit{b307}, \textit{b294}, and \textit{b308}, marked by the vertical grey bars as well as text) are systematically underestimated
  due to the photometric blending in the 2MASS catalog.
  \textit{Bottom:} The seeing measured on the $K_S$-band observations. The changes in seeing and zero-points are not
  correlated.}
\label{fig:vista_zp}
\end{figure*}

CASU's calibration pipeline calculates the zero-points of individual pawprints as the mean
difference between the instrumental magnitudes of point sources and their converted 2MASS counterparts.
In fields of high stellar density, several VISTA sources can be blended into one 2MASS source. 
The positional cross-matching procedure of CASU does not take this effect properly into account by 
allowing loose tolerances, leading to the photometric zero-point bias revealed here. 
Fig.~\ref{fig:vista_zp} illustrates the spatial dependence of this bias, showing the 
change in magnitude zero-points as measured by the CASU pipeline during $\sim$4 hours of VISTA 
observations of VVV fields. For comparison, the variation in the atmospheric seeing measured on the $K_S$-band observations
is also shown. These two quantities are clearly not correlated with each other.
During this particular night, VISTA was acquiring multi-band observations for
the VVV survey, hence the availability of all three of the bands $J$, $H$, and
$K_S$. Near the beginning and the end of this night, images were taken in uncrowded 
fields far from the Galactic plane, while during the
middle of the night, the densest VVV tiles were observed.
Besides low stellar densities, high extinction can also reduce the size of this
bias by diminishing the detection rate of faint 2MASS sources.
It should be noted that quite possibly other effects contribute to these (apparent) biases:
on the one hand, it is not straightforward to disentangle sky transparency changes during the night from other
photometric biases.
On the other hand, stellar magnitudes converted to the VISTA version 1.3 system from 2MASS depend on the
individually estimated extinction values (especially in the $JH$ bands, see Eqs.~\ref{eq:jv13}--\ref{eq:kv13}),
which is also estimated differently for the two versions,
therefore they are not as robust as those based on the more recent v1.5 conversion formulae.}

CASU also provide photometric catalogs for tile images, resulting from the combinations of six pawprints 
acquired at the same epoch. \citet{2018MNRAS.474.5459G}
describe a grouting process for the correction of tile-based photometry, whereby they take into account the 
photometric zero-point offsets between individual pawprints of a tile, among other effects. 
They attribute these offsets to variations in the PSF between successive pawprint observations.
However, we stress that 
varying levels of zero-point calibration bias due to blending also come into play in generating such offsets 
within a single observational epoch.
Each pawprint of a tile covers a slightly different subregion of a field due to their pointing offsets. 
If the distribution of interstellar extinction significantly differs between these subregions, then the amount of crowding, 
and hence the blending in the 2MASS catalog, will vary between pawprints.
As blending biases the pawprint zero-points, varying levels of blending between VISTA pawprints increase the scatter
of zero-points measured by the CASU pipeline, providing an additional source of pawprint-to-pawprint variation,
measured and partially corrected in the grouting process.
We hypothesize that this effect
can be seen in
Fig.~\ref{fig:vista_zp} as an increased scatter in the pawprint zero-points for fields with high stellar density;
however, the less robust calibration of version 1.3 of the VISTA photometry probably also contributes
to the high scatter observed in Fig.~\ref{fig:vista_zp} for crowded tiles.

\section{Recalibration of the $JHK_S$ observations in the VVV survey} \label{sec:recalibration}

As the photometric biases detailed in Sects.~\ref{sec:h_band} and \ref{sec:blending}
mostly manifest themselves in the zero-points of the VIRCAM detectors and the VISTA pawprint
observations, respectively, they can mostly be corrected for by a revision of the per-detector zero-points.

\begin{figure*}[t!]
  \centering
  \includegraphics[scale=0.46]{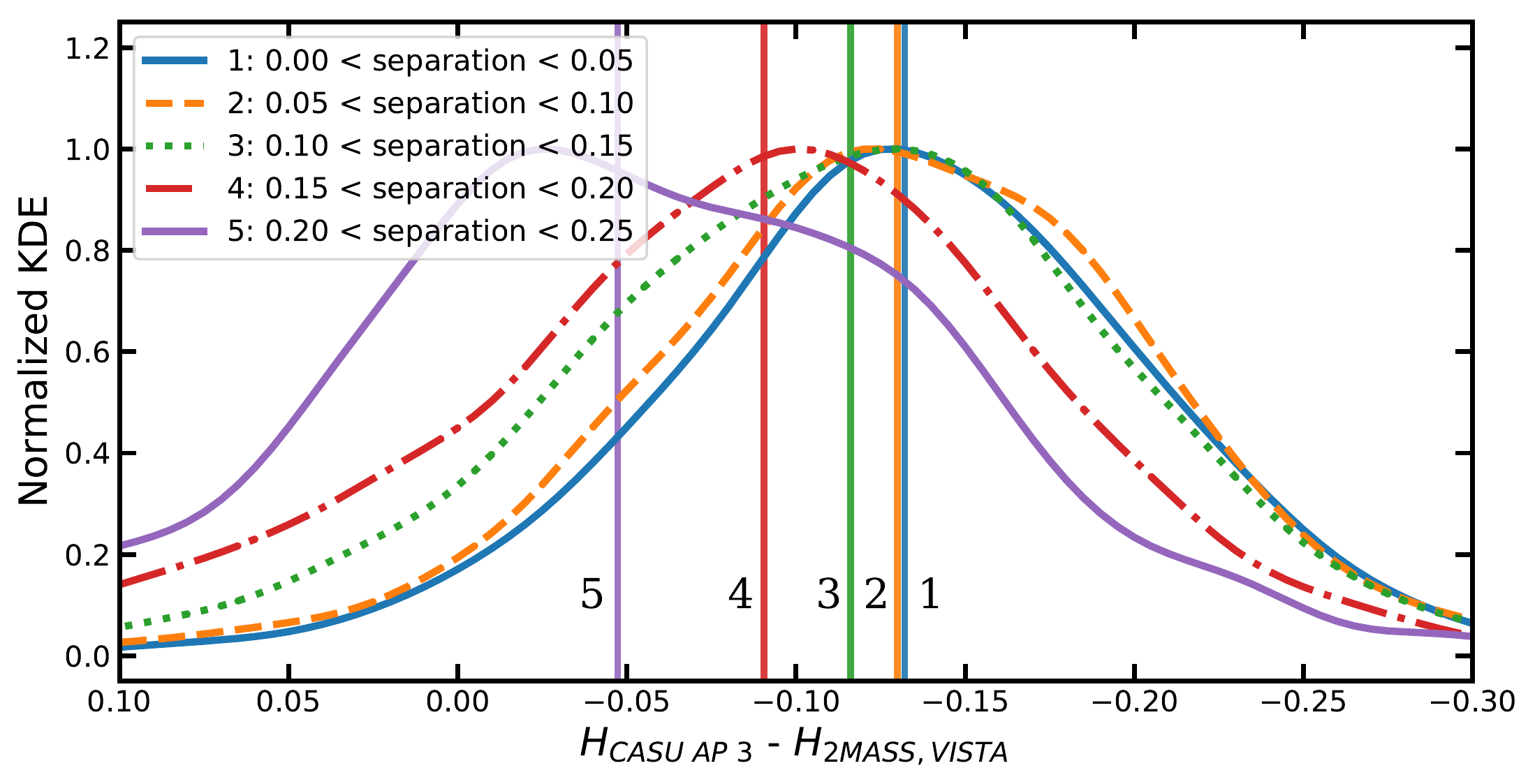}
  \caption[Dependency of magnitude offsets on the cross-match distance]{Kernel density estimates (KDEs) of
  magnitude differences in different cross-match angular separation bins in the $H$-band in VVV tile \textit{b307}
  (see middle panel of Fig.~\ref{fig:b307_jh}, for stars fainter than 12.5 magnitudes in the 2MASS catalog).
  Different curves correspond to different matching distances, as indicated in the inset (in arcsec).
  The KDEs are normalized to their individual maxima.
  Vertical bars mark the median values of different bins.
  Note the progressively fainter magnitudes measured by VISTA with increasing cross-match radius.}
\label{fig:b307_kde}
\end{figure*}

The zero-point calculation procedure of \citet{2018MNRAS.474.5459G} allows a relatively large,
1 arcsecond cross-match radius of tolerance between positions in the 2MASS catalog
and the
VVV source catalogs.
The color scale in Fig.~\ref{fig:b307_jh} shows the angular separation between the VVV sources
and their cross-matched 2MASS equivalents. Most notably, the 2MASS stellar sources lying above the fiducial lines
(hence probably being blends in the 2MASS catalog) generally have higher separations than those belonging
to the main locus of unblended stars. Naturally, as many sources are blended in the 2MASS Point Source
Catalog towards fields with high stellar density, not only their magnitudes, but also their measured positions
are affected by the blending, which explains the increased positional errors.

The dependence of the bias on the cross-match radius is further examined in Fig.~\ref{fig:b307_kde} for the
same $H$-band pawprint of the VVV field \textit{b307}, shown previously on the
middle panel of Fig.~\ref{fig:b307_jh}.
It can be seen that for higher angular separations, the VVV magnitudes become
progressively fainter due to the increasing fraction of blending in
the 2MASS catalog with increasing tolerance in the cross-matching. With a tolerance of 0.1 arcseconds,
the effect of blending becomes minimal, as shown by the median values of the samples, marked by vertical bars.

We developed a method for correcting the magnitude zero-points of individual detectors, 
taking into account the specific properties of the VVV survey's observations. The calibration process described by
\citet{2018MNRAS.474.5459G} calculates zero-points on the pawprint level. The principal reason for
this is to guarantee that enough 2MASS stars are available for the robust determination of zero-points,
even in high Galactic latitude fields. As VVV fields are generally rich in stars, we calculate the zero-point corrections
for each detector chip separately in each pawprint of each epoch for the five smallest apertures.
The main points of this correction method are the following:
\begin{itemize}
  \item The 2MASS Point Source Catalog converted to the VISTA version~1.5 system (Sect.~\ref{sec:vvv_2mass}) is used as
  the source catalog for the (re)calibration of VVV observations.
  As in some cases non-linearity affects VISTA observations down to 12.5 magnitudes (see
  middle panel of Fig.~\ref{fig:b307_jh}),
  only observations fainter than this limit are used.
  \item Unified source catalogs of each VVV field are created using the procedure of \citet{2018ApJ...857...54D}.
  They are cross-matched with the converted 2MASS catalog with a tolerance of 0.15 arcseconds.
  The small cross-match radius helps in diminishing the effect of the photometric bias due to the different 
  blending characteristics between the 2MASS and VISTA observations
  (see Sect.~\ref{sec:blending}). Although this effect becomes non-marginal for cross-match radii larger than 0.1 arcseconds,
  it remains fairly small until 0.15 arcseconds (see Fig.~\ref{fig:b307_kde}), and the increased source counts help in deriving a more robust zero-point
  correction in tiles with lower stellar densities.
  \item The zero-point correction is calculated for each detector and each aperture of each pawprint observation as the
  median difference between the magnitudes of stellar sources and their corresponding converted 2MASS magnitudes.
  The median difference is preferred over the mean, as it is a more robust statistic in the presence of outliers
  (for example, variable stars).
  \item Individual light curves are corrected for the VISTA zero-point biases by subtracting the appropriate
  zero-point corrections from their original CASU magnitudes, on a frame-by-frame basis.
\end{itemize}

\begin{table}[t!]
  \caption{Individual photometric zero-point correction factors for the version 1.3 photometry,
  while also converting them on the v1.5 scale, calculated for each observation, detector, and aperture, in the VVV survey.
  This table is available in its entirety in machine-readable form as Electronic Supplementary Material of the journal.}\label{tab:corr}
	\centering
  \resizebox{1\textwidth}{!}{\begin{tabular}{c@{\hspace{0.2cm}}c@{\hspace{0.2cm}}c@{\hspace{0.2cm}}c@{\hspace{0.2cm}}c|
    c@{\hspace{0.2cm}}c@{\hspace{0.2cm}}c@{\hspace{0.2cm}}c@{\hspace{0.2cm}}c|
    c@{\hspace{0.2cm}}c@{\hspace{0.2cm}}c@{\hspace{0.2cm}}c@{\hspace{0.2cm}}c|c}
         &                  &    &  &        & & \multicolumn{3}{c}{Aperture$^\mathrm{c}$ } & &  & \multicolumn{3}{c}{$\sigma_\mathrm{tot.}$$^\mathrm{d}$ } & &\\
    \hline
    Field & OBS$^\mathrm{a}$  & Band  & Chip & ZPT$^\mathrm{b}$ &
    1 & 2 & 3 & 4 & 5 &
    1 & 2 & 3 & 4 & 5 &N$_\mathrm{stars}$$^\mathrm{e}$\\
    \hline
    d001 & v20100314\_00245 & $J$ & 1 & 23.702 & -0.011 & -0.012 & -0.011 & -0.013 & -0.017  & 0.053 & 0.046 & 0.043 & 0.044 & 0.052 &    585 \\
    d001 & v20100314\_00245 & $J$ & 2 & 23.699 & +0.008 & +0.007 & +0.006 & +0.002 & -0.003  & 0.055 & 0.055 & 0.051 & 0.050 & 0.052 &    834 \\
    $\vdots$  &$\vdots$  &$\vdots$  &$\vdots$  & $\vdots$ &
    $\vdots$  &$\vdots$  &$\vdots$  &$\vdots$  & $\vdots$ &
    $\vdots$  &$\vdots$  &$\vdots$  &$\vdots$  &$\vdots$  &$\vdots$\\
    d001 & v20100314\_00221 & $H$ & 1 & 23.848 & +0.009 & +0.014 & +0.014 & +0.013 & +0.007 & 0.061 & 0.056 & 0.048 & 0.048 & 0.057 &  497 \\
    d001 & v20100314\_00221 & $H$ & 1 & 23.850 & +0.031 & +0.033 & +0.031 & +0.028 & +0.024 & 0.060 & 0.052 & 0.050 & 0.049 & 0.056  & 787 \\
    $\vdots$  &$\vdots$  &$\vdots$  &$\vdots$  & $\vdots$ &
    $\vdots$  &$\vdots$  &$\vdots$  &$\vdots$  & $\vdots$ &
    $\vdots$  &$\vdots$  &$\vdots$  &$\vdots$  &$\vdots$  &$\vdots$\\
    d001 & v20100129\_00205 & $K_S$&1 & 22.913 & +0.009 & +0.012 & +0.013 & +0.015 & +0.021 & 0.076 & 0.063 & 0.062 & 0.058 & 0.066 & 470 \\
    d001 & v20100129\_00205 & $K_S$&2 & 22.907 & +0.021 & +0.024 & +0.029 & +0.032 & +0.039 & 0.068 & 0.063 & 0.059 & 0.060 & 0.063 & 651 \\
    $\vdots$  &$\vdots$  &$\vdots$  &$\vdots$  & $\vdots$ &
    $\vdots$  &$\vdots$  &$\vdots$  &$\vdots$  & $\vdots$ &
    $\vdots$  &$\vdots$  &$\vdots$  &$\vdots$  &$\vdots$  &$\vdots$\\
    \hline
    \multicolumn{16}{l}{\footnotesize $^{\mathrm{a}}$ Identifier of the VISTA pawprint.}  \\
    \multicolumn{16}{l}{\footnotesize $^{\mathrm{b}}$ Value of the magnitude zero-point header keyword MAGZPT in the CASU v1.3 files.}  \\
    \multicolumn{16}{l}{\footnotesize $^{\mathrm{c}}$ Magnitude zero-point corrections for the five smallest apertures, calculated from
    the median difference.}  \\
    \multicolumn{16}{l}{\footnotesize $^{\mathrm{d}}$ Standard deviation calculated from the median absolute deviation (in mag units), of
    the VISTA--2MASS measurements} \\
    \multicolumn{16}{l}{\footnotesize (i.e., MAD$\times1.4826$). The formal error of the calculated magnitude offsets should be
    taken as the value given here, divided}  \\
    \multicolumn{16}{l}{\footnotesize by the square root of the number of stars used to estimate the offset.}  \\
    \multicolumn{16}{l}{\footnotesize $^{\mathrm{e}}$ Number of VISTA--2MASS stars used to calculate the photometric offsets.}  \\
	\end{tabular}}
\end{table}

The format of the correction offsets, calculated for every combination of pawprint observations, VIRCAM detectors, and the five smallest apertures
provided by CASU, for the $JHK_S$ obserations in the VVV survey, are shown in Table~\ref{tab:corr}.

\section{Effects of the recalibration}

\begin{figure*}[]
  \centering  
\subfigure{
  \includegraphics[width=\textwidth]{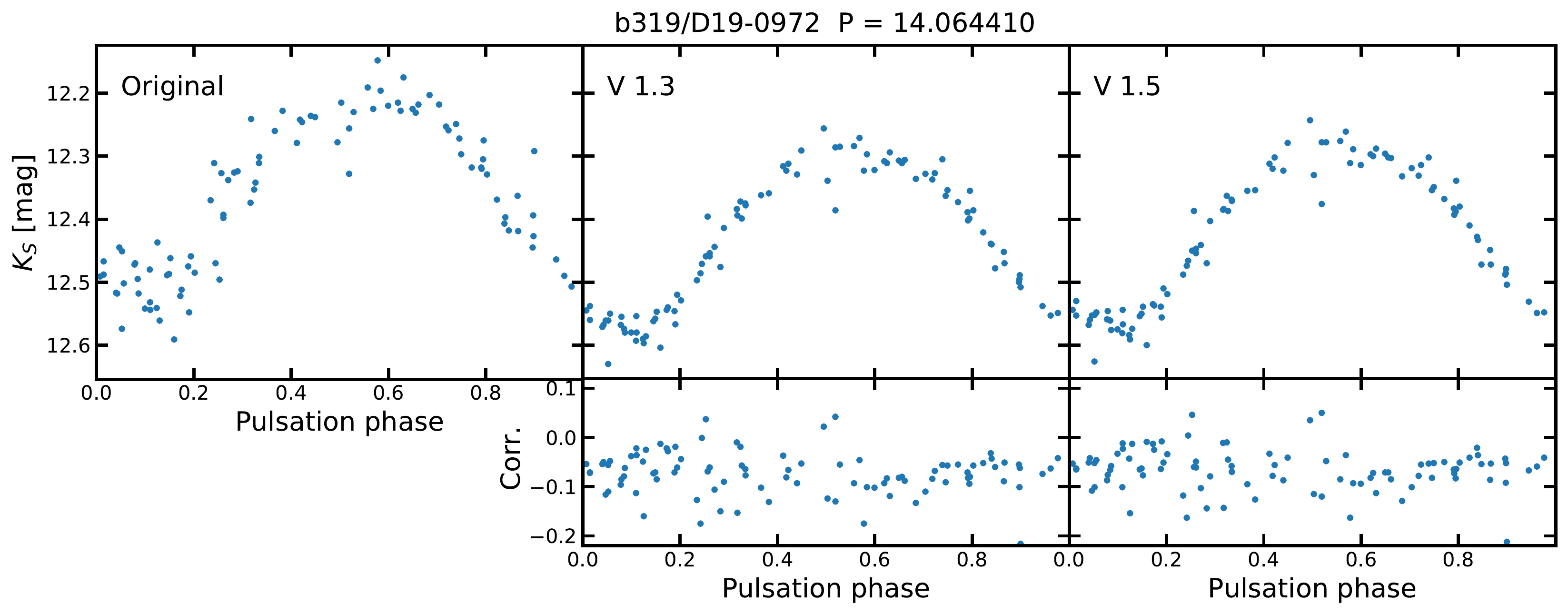}}%
\vskip-0.3cm
\qquad
\subfigure{
  \includegraphics[width=\textwidth]{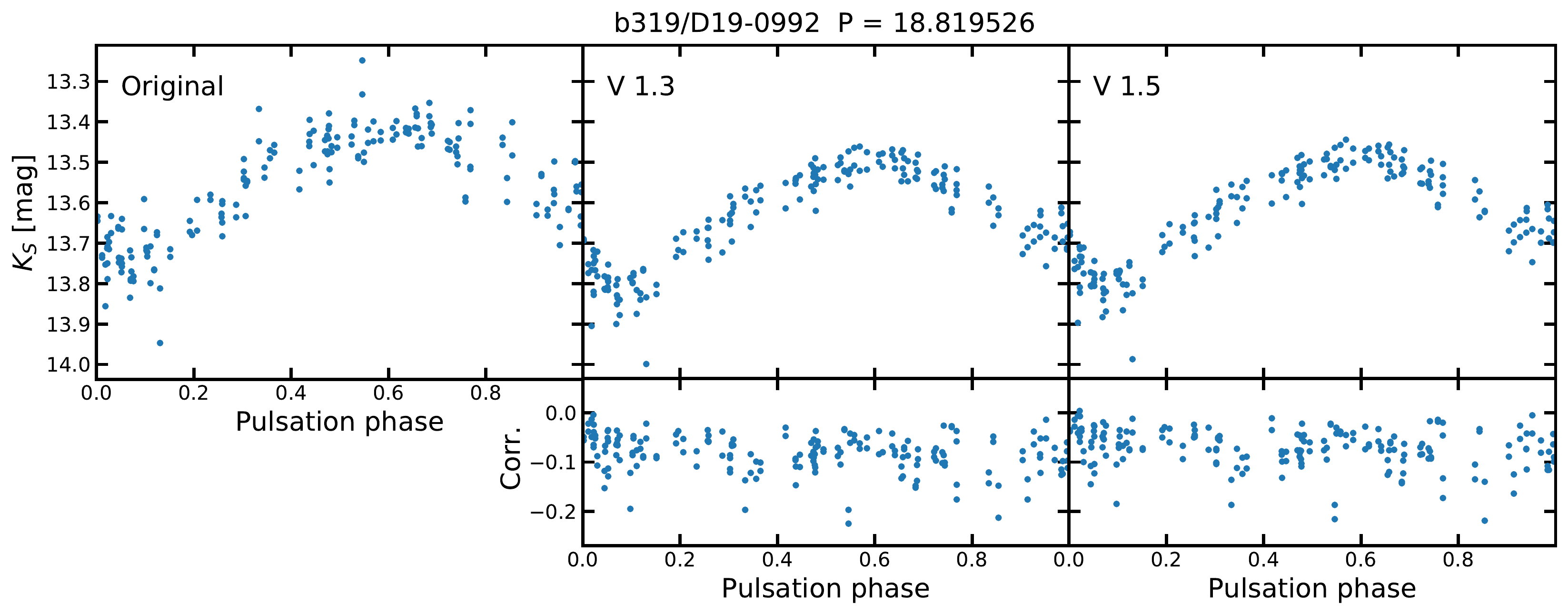}}%
\qquad
\subfigure{
  \includegraphics[width=\textwidth]{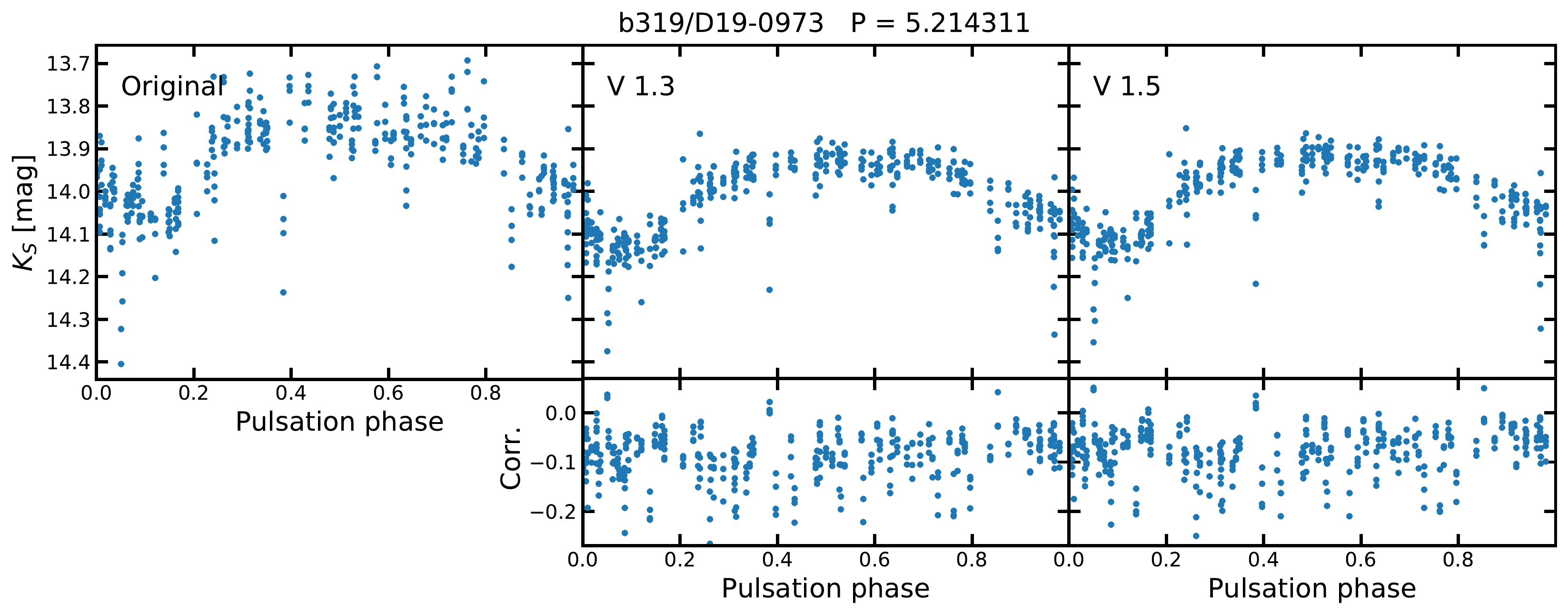}}%
\qquad
\subfigure{
  \includegraphics[width=\textwidth]{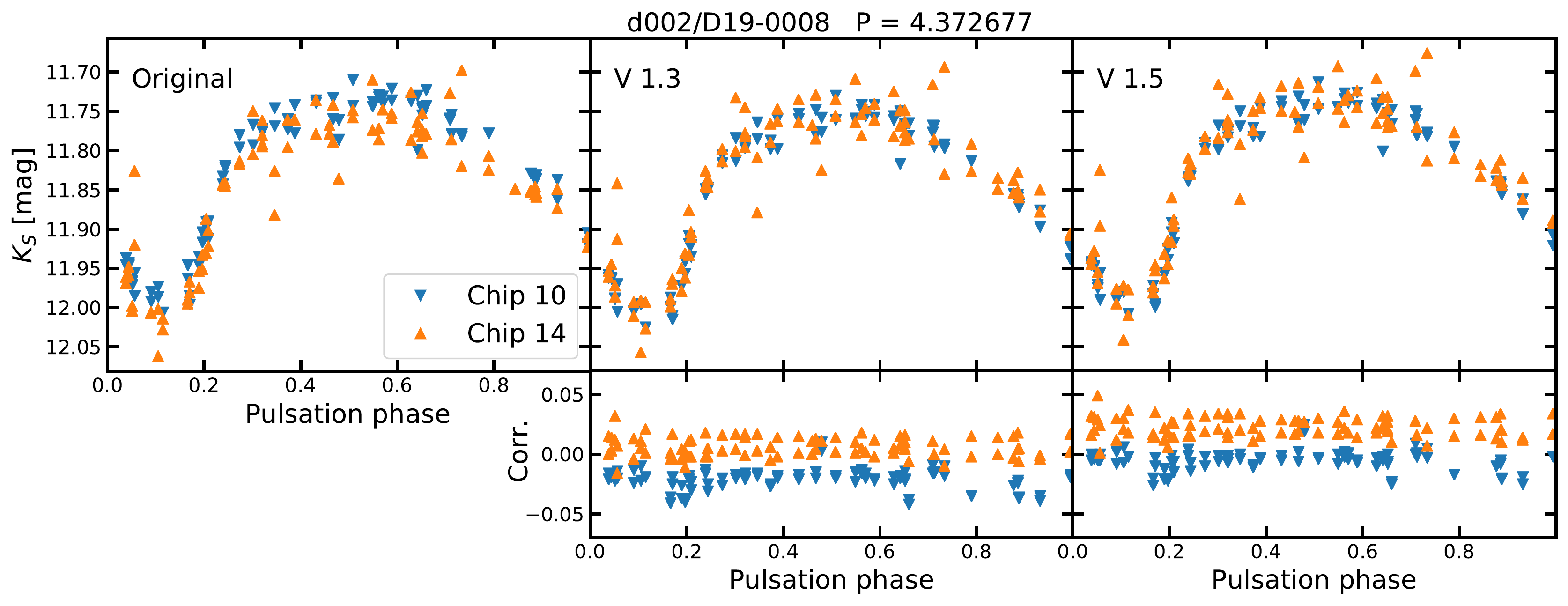}}%
  \caption{$K_S$-band light curves of Cepheids,
  before and after correction for the zero-point bias with the method described in Sect.~\ref{sec:recalibration}.
  \textit{Left panels:} The original phase-folded light curves.
  \textit{Middle panels:} The phase-folded light curves with the zero-point corrections applied in the VISTA v1.3 system.
  The lower panels show the individual correction factors.
  \textit{Right panels:} Same as the middle panels, but with the zero-point corrections appropriate for the
  VISTA v1.5 system applied (Table~\ref{tab:corr}).
  Note that the two Cepheids in the middle panels have observations from more than one pawprints from their tile,
  while the last Cepheid has VVV observations from two different VIRCAM detector chips.
  Above the panels belonging to each of the Cepheids, their VVV tile number, identification in the Cepheid catalog
  of \citet{2019ApJ...883...58D}, as well as periods in days are given.
  }
    \label{fig:cep_corr}
    \end{figure*}

After calculating the photometric zero-point corrections as described in Sect.~\ref{sec:recalibration}, the
full effect of the zero-point biases described in Sects.~\ref{sec:h_band} and \ref{sec:blending}
can be evaluated. 
However, in order to also assess the effect of the change of the conversion equations between
versions 1.3 and 1.5 of VISTA photometric system and 2MASS photometry, we have calculated the correction
offsets for the former version as well, allowing its direct estimation.
Fig.~\ref{fig:cep_corr} illustrates the effect of our recalibration 
on the light curves of four Cepheid variables. In the case of
three Cepheids from the moderately crowded tile \textit{b319}, their mean brightnesses in the $K_S$ band have
decreased by about 0.1 magnitudes. Furthermore, the scatter in their light curves decreased by about
one order of magnitude.
The fourth Cepheid illustrates the mostly constant photometric offsets between different chips in
the case of a VVV disk tile, which also contributes to the light curve scatter. The fact that the
quality of the light curves has improved so drastically validates the choice of the chip-based
recalibration of the photometric zero-points.
As for the effect of staying in the VISTA v1.3 or changing to the v1.5 system,
for individual stellar light curves,
the only obvious difference between them are the slightly higher zero points for the latter.

The spatial distribution of the magnitude offsets is affected by the pawprint pattern, as well as by
the apparent temporal variation of the zero-point bias, most probably due to changes in the
observing conditions and variations in telescope pointing.
The decreased scatter of the light curves shown in Fig.~\ref{fig:cep_corr} is a result of correcting for these effects.
However, it is also necessary to visualize the effect of the zero-point bias on the VISTA photometry over
small spatial scales.

\begin{figure*}[]
  \centering
  \includegraphics[width=0.9\textwidth]{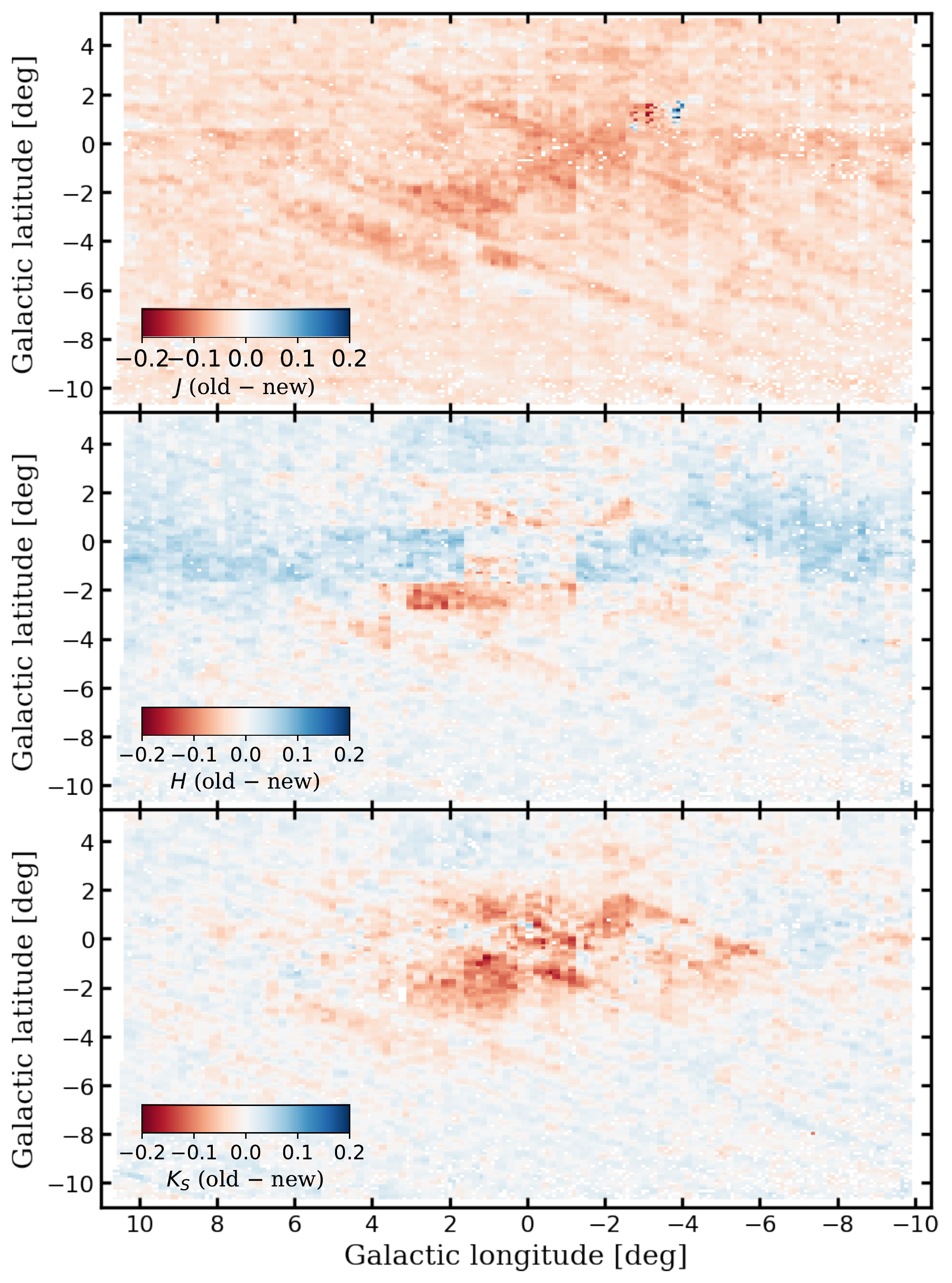}
  \caption{Top to bottom: $J$, $H$, and $K_S$-band maps of the median photometric
  correction factors (Table~\ref{tab:corr})
  over the complete VVV survey bulge area.
  Note that in the outer, no-crowded bulge regions the size of the correction factors
  are virtually equivalent to the average zero-point difference between versions 1.3 and 1.5
  of VISTA photometry \citep{2018MNRAS.474.5459G}.}
  \label{fig:bul_mag_diff}
\end{figure*}

We have constructed a uniform grid in Galactic coordinates, with gridpoints separated by $0.1^\circ$ in
both longitude and latitude, covering the entire bulge area of the VVV survey.
For each of the gridpoints, all sources within a cross-match radius of $5''$ were identified in the unified
source catalogs (Sect.~\ref{sec:recalibration}).
These allowed us to determine which of the VIRCAM chips fell onto each of the
gridpoints, along with the timings of the observations.
In turn, this information allowed the identification of the magnitude offset corrections
(calculated in Sect.~\ref{sec:recalibration}) which contribute to the magnitude differences at each of these
coordinates.
The spatial distribution of the median values of the corrections in the $J$, $H$, and $K_S$ bands
across the bulge area are shown in Fig.~\ref{fig:bul_mag_diff}.
Likewise, the biases in their color indices are shown in Fig.~\ref{fig:bul_color_diff}.
In the outer regions of the bulge,
the magnitude and color differences are minimal, due to the minimal blending of VISTA calibrator stars in the 2MASS catalog.
The small systematic differences in each band, seen towards the outer regions of the Galactic bulge,
are caused by the systematic difference between the VISTA version 1.3 and 1.5 photometric systems.
In contrast to this, in the tiles closest to the Galactic plane, the different handling of the color excess caused
by interstellar reddening in the two versions of the VISTA photometry causes an additional systematic offset: $\mathrm{E}(B-V)_{v1.3}$
is limited to a maximum value of 10, while $\mathrm{E}(B-V)_{v1.5}$ has a maximum value of $\sim4$ (the exact value depends on the
observed $J-K$ color of the particular 2MASS source). This effect can be most clearly seen on the top panel of Fig.~\ref{fig:bul_color_diff}.
Besides these offsets, toward regions of high stellar density, such as Baade's Window at $l,b \sim 2.5\,\mathrm{deg},-2.5\,\mathrm{deg}$, the
difference increases drastically. It has to be noted that this difference behaves markedly differently from
one filter to the next, most probably due to the different intrinsic colors and amount of extinction that each
star in the converted 2MASS catalog (Sect.~\ref{sec:vvv_2mass}) is subject to. The diagonal streaks, mostly 
seen in the $J$ band across the southern side of the bulge, are artifacts following the
2MASS scanning pattern and point toward zero-point offsets on the order of 0.01\,mag in those regions.
These artifacts appear because while we computed the photometric zero-point offsets individually for each chip,
in the original CASU photometry, the zero points were determined by per pawprint, thus they are averaged over these areas.

In order to evaluate the effect of changing the 2MASS--VISTA conversion formulae between versions 1.3 and 1.5,
the construction of photometric offset maps was repeated with the correction factors calculated for the former system.
The resulting maps showed the same structure as seen in Figs.~\ref{fig:bul_mag_diff} and \ref{fig:bul_color_diff},
therefore they are not repeated here. In is worth noting that due to the offset between the magnitude scales 
between the two systems (Equations~C9--C11 of \citealt{2018MNRAS.474.5459G}), the calculated correction
factors become virtually zero in the outer regions of the bulge in the case of version 1.3 of VISTA photometry.

\section{Conclusions and recommendations for the future usage of VISTA observations}

\begin{figure*}[]
  \centering
  \includegraphics[width=0.9\textwidth]{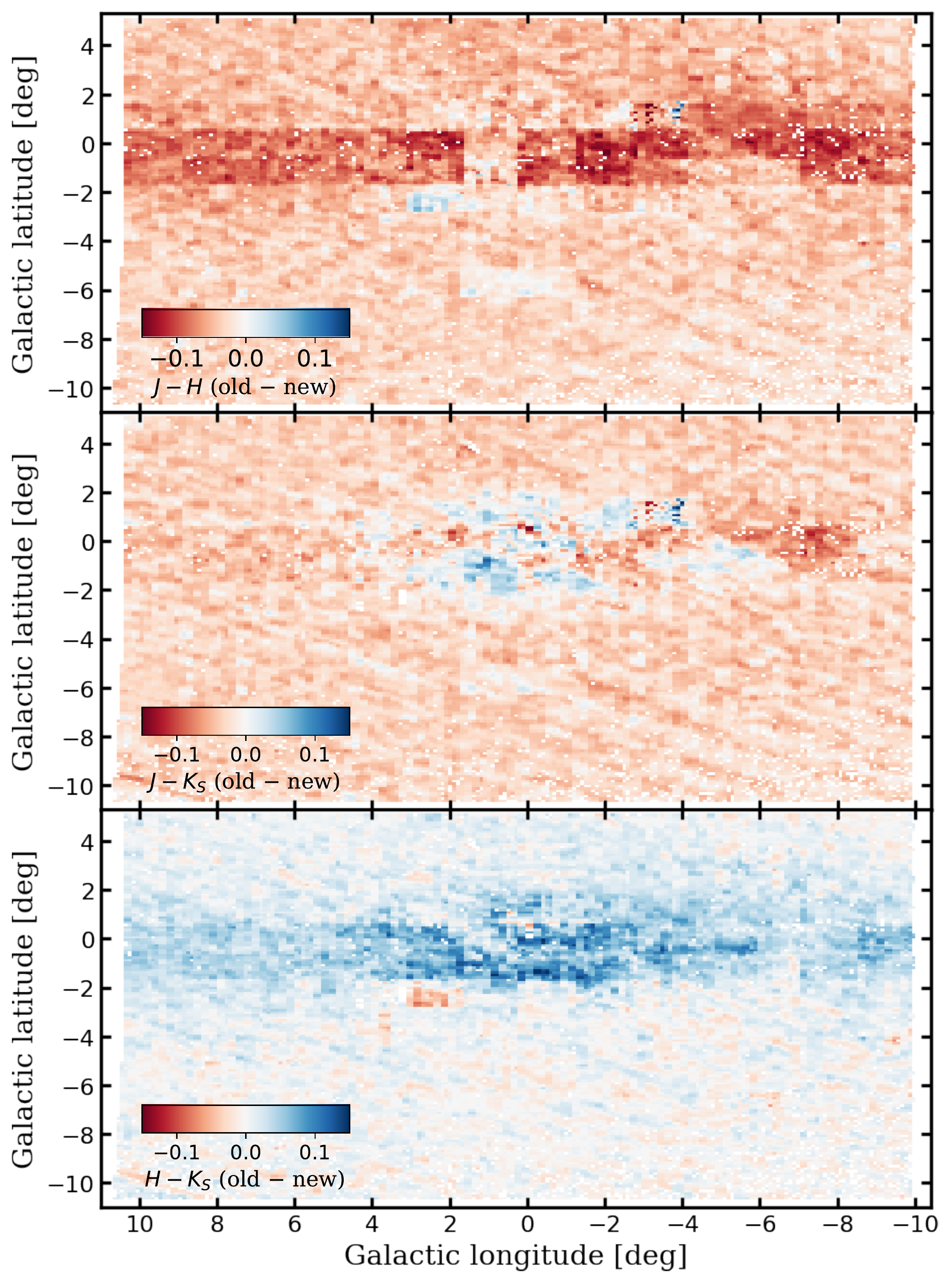}
  \caption{As in Fig.~\ref{fig:bul_mag_diff}, but the three panels from top to bottom show the 
  median $J-H$, $J-K_s$ and $H-K_s$ color differences, respectively.}
  \label{fig:bul_color_diff}
\end{figure*}

We have shown that the standard photometric source catalogs of the VVV survey produced by the VDFS 
suffer from space- and time-varying zero-point biases.
For this reason, some of the results based on VVV (and, more generally, VISTA) observations
will probably need to be revised, in order to be placed on a firmer footing.
For example, studies of the extinction law would likely benefit from a revision,
since both the derived apparent magnitudes and colors vary
towards different regions of the Galactic bulge (Figs.~\ref{fig:bul_mag_diff} and \ref{fig:bul_color_diff}). 
For example, \citet{2016MNRAS.456.2692N} studied the spatial variation
of the extinction law based on combined VVV $JHK_S$ and OGLE-IV $VI$ photometry by tracing the changes
in the color excesses of the red clump stars across the bulge. Although they measured stellar magnitudes
by a custom point-spread function (PSF) fitting photometry, their zero-point calibration is based on the CASU photometric
catalogs, and thus their data likely inherit the spatial zero-point bias structure shown in
Fig.~\ref{fig:bul_mag_diff}. \citet{2016MNRAS.456.2692N} found pairs of sight-lines where
the respective red clumps occupy the same position on the $K_S$ vs $V-I$ color-magnitude diagram, 
yet have very different $J-K_S$ colors (their Figs.~5 and 6). Based in part on this, they concluded that the
extinction curve has a non-standard shape towards the Galactic bulge.
As the compared fields are not on the same VVV pawprint or even tile, their $J$ and $K_S$ magnitude 
zero-points, and thus every color index derived from these quantities is
likely affected by the zero-point bias in different ways.
Studies of the nature, shape, and spatial variations of the extinction law would thus likely benefit
from a deeper reexamination of the VVV photometry.

Another area profiting immensely from the improved photometry is variable star studies. A cautionary
example is that of our own search for Classical Cepheids in VVV photometry. In \citet{2015ApJ...821L..29D},
we have claimed the existence of Classical Cepheids in a thin disk within the volume of the Galactic bulge, based on VVV data.\
This claim was challanged by \citet{2016MNRAS.462..414M}, based on their own near-infrared observations.
As pointed out in Section~6 of \citet{2019ApJ...883...58D}, the tension was partly caused by the $\sim0.1$\,mag
overestimation of the $E(H-K_\mathrm{S})$ color excess in the uncorrected VISTA magnitudes, which propagated into
the calculated absolute extinctions, and hence distances of the variables. The VISTA photometry, after applying the correction
factors given in Table~\ref{tab:corr}, gives consistent color excess estimates for the Cepheids common
between our work and that of \citet{2016MNRAS.462..414M}.

In this sense, the relatively simple recalibration procedure described in Sect.~\ref{sec:recalibration}, and provided
in Table~\ref{tab:corr}, should serve only as a
temporary measure in future studies utilizing VISTA observations.
In the case of the anomalous $H$-band measurements affected by the high sky background, we recommend that they are
discarded in future studies. However, in some cases they might be the only $H$-band measurements available for some
sources. Therefore, zero-point corrections are also included for these data in Table~\ref{tab:corr}, with the caveat that the
measurements will have high errors due to the non-linearity of the detectors (especially chips 5 and 6) for such
observations.
The selection of non-blended 2MASS stars
based on their cross-match radius that we adopted is a very preliminary approach, as it still leaves a considerable
fraction of blended stars in the sample for certain fields that is used for computing the zero-point corrections.
Ideally, each and every 2MASS star used in the zero-point calibration of VISTA observations should be revised
and potentially discarded if the VISTA observations reveal close companions within the 2MASS PSF.

Particularly in the most crowded VVV fields, PSF photometry has already proven its great value in
generating deep color-magnitude diagrams and enabling time-series studies, even of faint sources
\citep[e.g.,][]{2018A&A...619A...4A,2018ApJ...863...79C,2019arXiv190412024B}.
For PSF photometry, the light-curve scatter caused by the varying offsets between individual
epochs (see the difference in scatter between corrected and uncorrected light-curves in Fig.~\ref{fig:cep_corr})
is expected to be minimized, if the data are calibrated to a single epoch of CASU aperture photometry.
However, the spatially varying zero-point offsets shown in Fig.~\ref{fig:bul_mag_diff} might be
expected to directly propagate into the zero points of the PSF photometry if said catalogs are used
to derive them (as commonly done).
As an alternative, a custom calibration based on the converted
2MASS photometry can be recommended for the zero-point calibration of PSF measurements,
together with implementing additional steps for rejecting blended stars in the 2MASS catalog. 
Towards the Galactic center, PSF photometry has the unique
power of resolving individual stellar sources in the VVV survey. The centermost tiles are generally
observed right after and/or before more outlying tiles. Therefore, the zero-points derived for
these flanking tiles could be adopted as the zero-points for the photometry of Galactic center tiles.

Alternatively, the repeated observations of tiles under photometric conditions
could be used in an
``\textit{Ubercal}''-like calibration scheme \citep{2012ApJ...756..158S,2013ApJS..205...20M},
allowing the absolute zero points of dense tiles to be calibrated internally to less dense tiles.
In the case of the VVV survey, in the $YZJH$ bands the number of repeated observations is very low,
and as seen in Sect.~\ref{sec:h_band}, observations were sometimes taken under poor observing conditions,
prohibiting their use for \textit{Ubercal}.
Therefore, most probably only the repeated $K_\mathrm{S}$-band observations of VVV would be suitable for this
calibration scheme for the complete survey area. Given the high amount of crowding present in the most densely
populated tiles of the VVV survey, this operation should attempted using PSF photometry, preferably.

\begin{acknowledgements}
We gratefully acknowledge the constructive comments provided by the referee, Carlos
Gonz\'alez-Fern\'andez, which helped to improve this work.
G.H. and M.C. gratefully acknowledge the support provided by FONDECYT through grant \#1171273;
by the Ministry for the Economy, Development, and Tourism's Millennium Science Initiative through
grant IC\,120009, awarded to the Millennium Insitute of Astrophysics (MAS); and by Proyecto Basal AFP-170002.
I.D. and E.K.G. were supported by Sonderforschungsbereich SFB 881 ``The Milky Way System'' (subproject A03)
of the German Research Foundation (DFG Project-ID 138713538).
The research leading to these results has received funding from the European Research Council (ERC) under the
European Union's Horizon 2020 research and innovation programme (grant agreement No~695099).

\end{acknowledgements}

\bibliographystyle{spbasic}


\end{document}